\documentclass[%
reprint,
superscriptaddress,
showpacs,
 amsmath,amssymb,
 aps,
prb,
]{revtex4-1}

\usepackage{graphicx,subfigure,bbm}
\usepackage{dcolumn}
\usepackage{bm}

\usepackage{amsmath}
\usepackage{amsthm}

\usepackage[english]{babel}
\usepackage[utf8]{inputenc}
\usepackage{graphicx}
\usepackage[colorinlistoftodos]{todonotes}
\usepackage{bbold }
\usepackage{physics}
\usepackage[utf8]{inputenc}
\usepackage{amssymb }
\usepackage{graphicx}
\DeclareGraphicsExtensions{.eps}
\usepackage{epstopdf}
\usepackage{latexsym}
\usepackage{amsfonts}
\usepackage{xcolor}
\usepackage{units}
\usepackage{bm}
\usepackage{verbatim}
\usepackage[colorlinks = true,
            linkcolor = red,
            urlcolor  = blue,
            citecolor = magenta,
            anchorcolor = red]{hyperref}
\usepackage{esint}
\usepackage{soul}
\usepackage{cancel}
\usepackage[normalem]{ulem}
\usepackage{empheq}
\usepackage{dsfont}
\DeclareSymbolFont{matha}{OML}{txmi}{m}{it}
\DeclareMathSymbol{\varv}{\mathord}{matha}{118}

\DeclareMathOperator{\arccosh}{arcCosh}

\newcommand{\sign}{\text{sign}}

\definecolor{GB-color}{RGB}{255,0,255}
\definecolor{AB-color}{RGB}{0,0,255}
\begin{document}


\title{Nonlocal thermoelectric engines in hybrid topological Josephson junctions}

\author{Gianmichele Blasi}
\email{gianmichele.blasi@sns.it} 
\affiliation{NEST, Scuola Normale Superiore and Istituto Nanoscienze-CNR, I-56126, Pisa, Italy}
\author{Fabio Taddei}
\affiliation{NEST, Scuola Normale Superiore and Istituto Nanoscienze-CNR, I-56126, Pisa, Italy}
\author{Liliana Arrachea}
\affiliation{International Center for Advanced Studies, ECyT-UNSAM, Campus Miguelete, 25 de Mayo y Francia, 1650 Buenos Aires, Argentina}
\author{Matteo Carrega}
\affiliation{SPIN-CNR, Via Dodecaneso 33, 16146 Genova, Italy}
    
\author{Alessandro Braggio}
\email{alessandro.braggio@nano.cnr.it} 
\affiliation{NEST, Scuola Normale Superiore and Istituto Nanoscienze-CNR, I-56126, Pisa, Italy}

\begin{abstract}
The thermoelectric performance of a topological Josephson nonlocal heat engine is thoroughly investigated. The nonlocal response is obtained by using a normal metal probe coupled with only one of the proximized helical edges in the middle of the junction. In this configuration, we investigate how the flux bias and the phase bias trigger the nonlocal thermoelectric effects under the application of a thermal difference between the superconducting terminals. Possible experimental nonidealities such as asymmetric proximized superconducting gaps are considered showing how the nonlocal response can be affected.
The interplay between Doppler-shift, which tends to close gaps, and Andreev interferometry, which affects particle-hole resonant transport, are clearly identified for different operating regimes. Finally, we discuss the power and the efficiency of the topological thermoelectric engine which reaches maximum power at maximal efficiency for a well coupled normal probe. We find quite high nonlocal Seebeck coefficient of the order of tenths of $\mu$V/K at a few kelvin, a signal that would be clearly detectable also against any spurious local effect even with moderate asymmetry of the gaps.    
\end{abstract}
\date{\today}
\maketitle


\section{Introduction}
\label{Introduction}
Topological Josephson junctions have been actively investigated in the recent 
years.\cite{Sacepe11,Veldhorst12,Hart14,Sochnikov15}
In particular, their potential to 
localize Majorana fermions\cite{Fu08,Fu09,Schrade15,Stehno16}
 could represent a novel platform for topological quantum computation.\cite{Virtanen18,Hedge20} 
Topological Josephson junctions are also a unique resource in the field of low-temperature 
thermal management\cite{Giazotto06,Partanen16,Fornieri17,Hwang20}, 
 which could play an important role for quantum technologies in general. New applications based on the proximized helical 
 edge states have been envisioned.\cite{Tkachov15,Sothmann16,Sothmann17,Bours18,Blasi19,Vischi19,Scharf20} 
For such cases a fundamental step is the capability to identify the helical nature of the edge states as result of the spin-momentum locking, determined by the topological protection.\cite{Moore09,Hasan10,Qi11,Ando13}
After the theoretical prediction\cite{Kane05,Bernevig06}, experimental evidence\cite{Konig07,Roth09,Brune12} have been shown on the existence of edge states in several systems but not yet on their helical nature.\cite{Shi19,Wu17,Jia17,Reis17,Li18,Liu20,Pribiag15}
For this purpose different strategies have been identified.\cite{Das11,Mani17}
Recently, we have realized that the helical nature determines an unique signature in the thermoelectrical properties of topological Josephson junctions.\cite{Blasi_2020_PRL,Blasi_2020_Andreev_Inteferometer} 

Thermoelectricity is in itself an important trend in material science\cite{Goldsmid10}, which found a renaissance in low-dimensional \cite{Dresselhaus07,Dubi11} and quantum-based\cite{Whitney14,Sothmann14} devices.
The thermoelectric response in superconducting systems is expected to be negligible due to the particle-hole (PH) symmetry which is enforced by superconducting correlations as clearly shown in the Bogoliubov-de Gennes (BdG) Hamiltonian.~\cite{DeGennes} Still, thermoelectric response of superconductors has a long history since the Ginzburg seminal work~\cite{Ginzburg44} and following literature~\cite{Galperin74,Shelly16}. There are various strategies aimed at inducing thermoelectricity in superconducting or proximized systems by explicitly breaking the PH symmetry by means of ferromagnetic correlations\cite{Machon13,Oazeta14,Kolenda17,Shapiro17,Keidel20} and nonlocal geometries\cite{Machon13,Mazza15,Heidrich19} or by using nonlinearities.\cite{Sanchez16,Pershoguba19,Marchegiani20,Marchegiani20b} 
Recently, several authors have discussed Andreev interferometers~\cite{Virtanen04,Jacquod10,Kalenkov17,Kalenkov20,Titov20} and nonlocal thermoelectricity in Cooper pair splitters~\cite{Hussein19,Kirsanov19}, which found experimental confirmations.~\cite{Chandrasekhar98,Jiang05,Petrashov03,Tan20} The application of thermal gradients to Josephson junctions has suggested also novel technologies\cite{Giazotto15,Hekkila18,Guarcello18,Guarcello19,Marchegiani20,Marchegiani20c,Marchegiani20d}, showing that the peculiar properties of topological Josephson junctions could also play an important role
in this perspective.\cite{Bours18,Bours19,Kamp19,Scharf20,Strambini20,Guarcello20,Gresta_Blasi_2021}
 
In this work we generalize the analysis of Refs.~\onlinecite{Blasi_2020_PRL} and \onlinecite{Blasi_2020_Andreev_Inteferometer}, where we discussed how the nonlocal thermoelectric response is intimately connected to the helical nature of edge states in a topological Josephson junction (TJJ). In these previous works we analyzed a three terminal structure where the 
TJJ, obtained by proximizing the two ends of a 2D topological insulator (TI) bar through superconducting electrodes, is contacted on one edge with a normal metal probe (N).\cite{Das11,Liu15,Hus17,Voigtlander18,Bours18,Bours19} In this configuration there is a nonlocal thermoelectric response when a temperature bias is applied between the two superconductors, which consists in the occurrence of a current flowing in the probe.
We note that in order to observe such nonlocal thermoelectric effect it is necessary to break the particle-hole (PH) symmetry at nonlocal level.\cite{Benenti17}
This can be done by introducing a magnetic field orthogonal to the plane of the TI, which induces the so called Doppler shift (DS) of the TI bands\cite{Blasi_2020_PRL}, or simply by applying\cite{Blasi_2020_Andreev_Inteferometer} a Josephson phase difference $\phi$, which may be generated by imposing a dissipationless current throughout the junction.
In Refs.~\onlinecite{Blasi_2020_PRL,Blasi_2020_Andreev_Inteferometer} we assumed equal proximized gaps in the two superconducting right/left ends and we concentrated only on the 
linear regime.
In this work we generalize to the case where we have different gaps, showing how the unique nonlocal signature survives in the 
asymmetric case.  At the same time we take the opportunity to analyze in details the nonlinear regime, discussing the thermodynamic performance of the nonlocal thermoelectric engine obtained in such configuration.

More specifically, in Sec.~\ref{The_System}, we present the three terminal setup and introduce the model Hamiltonian. By using the scattering approach, we discuss how the dissipative currents can be computed. In Sec.~\ref{Linear_response_regime} we investigate the nonlocal Onsager coefficients in the linear-response regime and discuss how the junction asymmetry affects the nonlocal thermoelectricity. In Sec.~\ref{Non-linear response regime} we study the performance quantifiers (electrical power, efficiency and nonlocal Seebeck coefficient) in the non-linear regime and in the presence of finite voltage bias and/or finite temperature difference between the two superconducting electrodes. In Sec.~\ref{Conclusions} we present the relevant conclusions concerning the nonlocal effects of the topological Josephson junctions and possible perspectives.

\section{System and Model}
\label{The_System}
\begin{figure}
\centering
    \includegraphics[width=.49\textwidth]{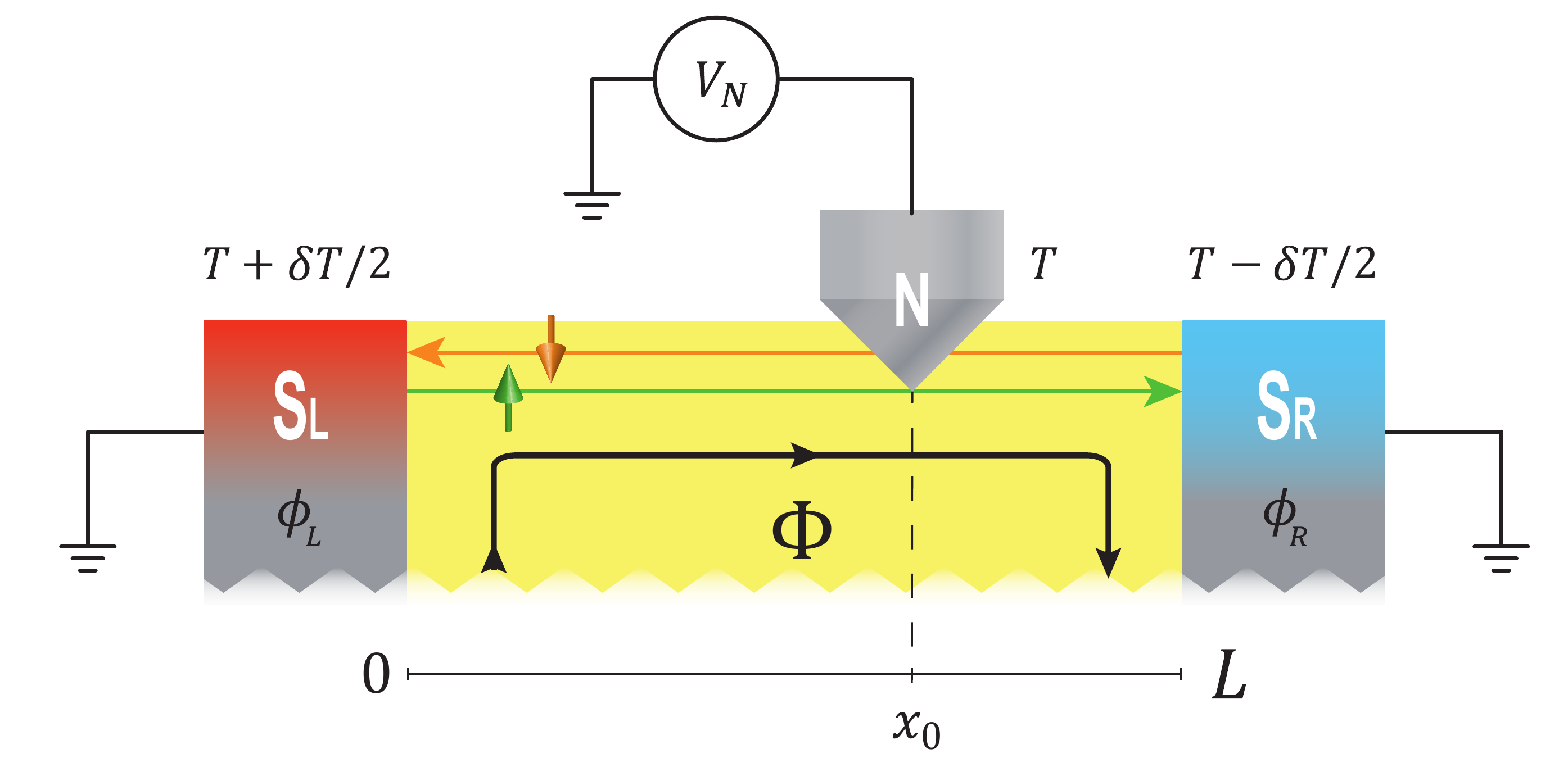}
    \caption{Sketch of the setup. A helical Kramers pair of edge states of a quantum spin Hall bar is put in contact with two superconductors at different temperatures $T_{L}=T+\delta T/2$ and $T_{R}=T-\delta T/2$. A bias voltage $V_N$ is applied to the normal-metal probe coupled to the edge at the point $x_0$ and kept at temperature $T_N=T$. $L$ is the length of the junction. The structure is threaded by a magnetic flux $\Phi$ which induces a Doppler shift in the edge states in addition to a Josephson phase difference $\phi\equiv \phi_{R}-\phi_{L}$ applied between the two superconductors. The green arrow depicts spin-$\uparrow$ right-moving quasi-particles, the orange arrow indicates spin-$\downarrow$ left-moving quasi-particles.} 
    \label{scheme_with_temperature_voltage} 
\end{figure}

We consider a TJJ which consists of two superconducting electrodes placed on top of a 2D TI at a distance $L$ (see Fig.~\ref{scheme_with_temperature_voltage}). 
The two electrodes induce superconducting correlations on the edge states via proximity effect~\cite{Tkachov15,Sothmann16}.
A normal-metal 
probe is contacted to one side of the junction as depicted in Fig.~\ref{scheme_with_temperature_voltage}, thus putting it in electrical contact with only one edge.
In the setup, a voltage bias $V_N$ may be applied between the probe N and the superconducting electrodes,  which are equipotential and grounded.~\footnote{Keeping the two superconductor equipotentials is important in order to \emph{not} induce time dependent Josephson effect.}
The two superconductors are kept at different temperatures in order to maintain a thermal bias $\delta T=T_{L}-T_{R}$.
A superconducting phase difference $\phi=\phi_{R}-\phi_{L}$ is also applied between the two.
We fix the temperature of the probe at the average temperature $T_{N}=(T_{L}+T_{R})/2=T$. As we will see below, this choice is the most convenient experimental one in order to extract the nonlocal signal.
The width of the TI strip is assumed to be large enough such that we can neglect the lower edge, i.e.~we focus only on the upper one.

The proximized system in the upper-edge can be described by the following Bogoliubov-de Gennes (BdG) Hamiltonian 
\begin{equation}
\label{My_Hamiltonian}
 {\cal H}=\mqty(H(x) & i\sigma_y\Delta (x) \\ -i\sigma_y\Delta (x)^*  &  -H(x)^*) ,
\end{equation}
expressed in the four-component Nambu basis $(\psi_{\uparrow},\psi_{\downarrow},\psi_{\uparrow}^{*},\psi_{\downarrow}^{*})^T$ with spin $\uparrow$ and $\downarrow$ collinear with the natural spin quantization axis of the TI edge pointing along the $z$-direction.
In Eq.~(\ref{My_Hamiltonian}) $H(x)=\varv_F\left(-i\hbar\partial_x+p_{DS}/2\right)\sigma_z-\mu\sigma_0+\Lambda(x)$ is the low-energy effective Hamiltonian of a 2D TI, with $-H(x)^*$ being its time-reversal partner.
Here $\varv_F$ indicates the Fermi velocity, $\mu$ is the chemical potential and $\sigma_i$ are the Pauli matrices.
The momentum $p_{DS}=(\pi \hbar/L)(\Phi/\Phi_0)$ represents the so-called DS contribution describing the gauge invariant shift of momentum induced by a small magnetic flux $\Phi$ threading the weak link, while $\Phi_0=h/2e$ is the magnetic flux quantum.~\cite{Tkachov15}
A contact potential $\Lambda(x)=\Lambda_L\delta(x)+\Lambda_R\delta(x-L/2) $ is also included at the junction boundaries.
$\Delta(x)$ is the superconducting order parameter which is assumed to obey rigid boundary conditions: $\Delta(x) = \Delta_{L}\Theta(-x)e^{i\phi_{\rm L}} + \Delta_{R}\Theta(x - L)e^{i\phi_{\rm R}}$, where $\Theta(x)$ is the step function and $\Delta_i$ (with $i={\rm L/R}$) is the proximity induced gap for the left and right superconducting regions. The two gaps can indeed be different for many reasons. For example, the two superconductors may be made of different materials\cite{Tinkham}, or the two superconducting films might have different  thicknesses\cite{Chubov69,Meserey71,Ilin04}, or the two S-TI interfaces might be made of different quality\cite{McMillan68,Maier12}.
Moreover, in realistic experimental conditions, the finite temperature difference in the nonlinear regime may induce the superconducting gaps to take different values on the two sides\cite{Tinkham}. In the latter case, in order to make realistic predictions in a wide temperature range, we need to include the self-consistent temperature dependence of the gaps.
\begin{figure*}[ht]
\centering
   \includegraphics[width=0.75\textwidth]{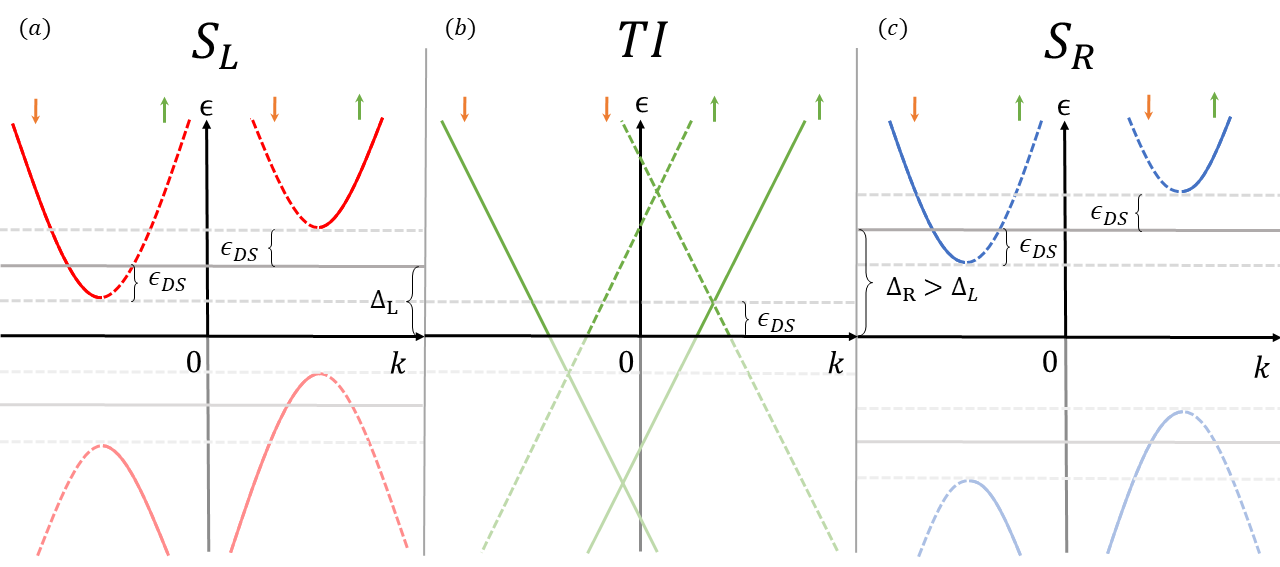}
    \caption{Dispersion curves for electron-like QPs (solid lines) and hole-like QPs (dashed lines) in the left [panel (a)] and right [panel (c)] proximized regions, and in the weak link TI region [panel (b)], for $0 <\epsilon_{DS}(\Phi) <\Delta_L$ (with $\Delta_R>\Delta_L$). Green arrows represent spin-$\uparrow$ right-moving QPs, while orange arrows spin-$\downarrow$ left-moving QPs.} 
   \label{dispersion} 
\end{figure*} 

The eigenspectrum of the BdG Hamiltonian relative to a homogeneously proximized 2D TI upper edge is given by $E^{i\chi }_{\pm}(k)=\left(\epsilon_{DS}(\Phi)+\chi \sqrt{(\hbar \varv_F k\mp \mu)^2+\Delta_i^2}\right)$.
Here the lower index $\pm$ labels the right/left parabola, $\chi =\pm$ indicates branches with positive/negative concavity, $i={\rm L,R}$ for the left and right superconducting regions and $\epsilon_{\rm DS}(\Phi)=\varv_F p_{DS}/2=(\varv_F h/4 L)(\Phi/\Phi_0)$ is the DS energy.
The effect of the DS on the dispersion curve is to shift the various branches vertically by an amount $\epsilon_{\rm DS}(\Phi)$, upwards or downwards, as shown in the example of Fig.~\ref{dispersion}, where we plot the dispersion curves for the three regions composing the TJJ.
As a consequence, a finite value of the magnetic flux $\Phi$ reduces the gap in the eigenspectrum, which eventually closes when $\abs{\epsilon_{\rm DS}(\Phi)}=\Delta_{\rm L/R}$ in the left/right superconducting regions of the 2D TI, respectively.
Clearly, when the values of $\Delta_{\rm L}$ and $\Delta_{\rm R}$ are different, such a gap closing in the left and right regions occurs at different values of the flux $\Phi$.

The eigenfunctions of the electron-like Bogoliubov quasiparticles (QPs) of the BdG Hamiltonian for both left/right superconducting regions can be written, in Nambu notation, as
\begin{eqnarray}
\Psi_{e_{+}}^{i,\chi }&=(\chi u_{i-} e^{i \frac{\phi_i}{2}},0,0,v_{i-} e^{-i \frac{\phi_i}{2}})^{T}e^{i k_{e_{+}}^{i\chi }x}\nonumber\\
\Psi_{e_{-}}^{i,\chi }&=(0,-\chi u_{i+} e^{i \frac{\phi_i}{2}},v_{i+} e^{-i \frac{\phi_i}{2}},0)^{T}e^{i k_{e_{-}}^{i\chi }x} ,
\label{Eigenfunctions}
\end{eqnarray}
where the energy-dependent coherence factors (spinor components) are
\begin{alignat}{2}
  u_{i\pm}  =\sqrt{\frac{\Delta_i}{2 \epsilon_{\pm}}} e^{\frac{1}{2}h_i(\epsilon_{\pm})};\quad&v_{i\pm}  = \sqrt{\frac{\Delta_i}{2 \epsilon_{\pm}}} e^{-\frac{1}{2} h_i(\epsilon_{\pm})}
  \label{u_v},
\end{alignat}
with $\epsilon_{\pm}=\epsilon\pm \epsilon_{DS}(\Phi)$ and $h_i(\epsilon_{\pm})= \arccosh{(\epsilon_{\pm}/\Delta_i)}$ for $\epsilon_{\pm}>\Delta_i$ and $h_i(\epsilon_{\pm})=i \arccos{(\epsilon_{\pm}/\Delta_i)}$ for $\epsilon_{\pm}<\Delta_i$.
The QP's momentum is $k_{e_\pm}^{i\chi }=\pm k_{F}(\chi  \sqrt{(\epsilon_{\mp}^2-\Delta_i^2)/\mu^2}+ 1)$, while the related group velocity is $\varv_{e_\pm}^{i\chi }=\hbar^{-1}|\partial_k E_{\pm}^{i\chi }|=\varv_F(u_{i\mp}^{2}-v_{i\mp}^{2})$. 
The eigenfunctions $\Psi_{h_{\pm}}^{i,\chi }$ relative to the hole-like Bogoliubov QPs can be obtained by replacing $(u_{i\pm},v_{i\pm})\rightarrow (v_{i\pm},u_{i\pm})$, $k_{e_\pm}^{i\chi }\rightarrow k_{h_\mp}^{i\chi }= k_{e_\pm}^{i,\bar{\chi} }$ (with $\bar{\chi}=-\chi$) and $\varv_{e_\pm}^{i\chi }\rightarrow \varv_{h_\mp}^{i\chi }=\varv_{e_\pm}^{i\chi }$    in the expressions of Eq.~(\ref{Eigenfunctions}).
Clearly, the limit $\Delta_i\to 0$ returns the standard 1D Dirac spectrum [shown in Fig.~\ref{dispersion}(b)] which characterizes the 2D TI (non-proximized) edge.

The normal-metal probe N -- which would model for instance a STM tip~\cite{Das11,Liu15,Hus17,Voigtlander18} -- is assumed to be directly contacted to the upper edge at the point $x_0$ (see Fig.~$\ref{scheme_with_temperature_voltage}$) and modelled by an energy- and spin-independent transmission amplitude $t$.

\subsection{Charge and Heat Currents}
\label{Charge_and_Heat_Currents}
In the following we will investigate only the dissipative currents flowing through the structure which can be obtained by using the Landauer-B\"uttiker scattering~\cite{Blanter00} formalism generalized in order to include superconductivity~\cite{Lambert98}.
More precisely, we are not interested in the dissipationless (Josephson) contribution to the charge current flowing in the superconducting electrodes, which has been already discussed elsewhere\cite{Tkachov13,Snelder13,Sochnikov15,Dolcini15,Kurter15,Marra16,Stehno16,Scharf20}.
The dissipative charge ($\gamma=-$) and heat ($\gamma=+$) currents exiting from lead $i={\rm L,R,N}$ can be written in this form
\begin{align}
\label{current_Lambert_general}
\begin{bmatrix}
J_i^{-} \\
J_i^{+} 
\end{bmatrix}
\!\!=&\frac{2}{h}\!\!\sum_{j,\alpha,\beta}\!\int_{0}^{\infty}\!\!\!\!\!\!\!d\epsilon 
\!\begin{bmatrix} q_i^{\alpha}(\epsilon)\\\epsilon-(q_i^{\alpha}(\epsilon)V_i-\alpha \mu_S)
\end{bmatrix}\! F^{\alpha\beta}_{ij}(\epsilon)
P_{i,j}^{\alpha,\beta}(\epsilon,\vec{\theta}) ,
\end{align}
where $\mu_S$ is the chemical potential of the
superconductors, which we take as a reference for all the energies, and $\alpha,\beta=+/-$ for electron-like and hole-like QPs, respectively. 
The function 
\begin{equation}
\label{F_fermi_difference}
F^{\alpha\beta}_{ij}=f_i^{\alpha}(\epsilon)-f_j^{\beta}(\epsilon)
\end{equation}
is a compact way to write
differences of the generalized Fermi functions $f^{\alpha}_i(\epsilon)=\{e^{\epsilon-\alpha(eV_i-\mu_S)/k_BT_i}+1\}^{-1}$,
where $T_i$ and $V_i$ are, respectively, the temperature and the voltage of lead $i$. As discussed earlier, in the following we will set $eV_{L}=eV_{R}\equiv \mu_S=0$.

In Eq.~(\ref{current_Lambert_general}) the most important physical quantities are the scattering coefficients $P_{i,j}^{\alpha,\beta}(\epsilon,\vec{\theta})$ which are calculated from the scattering matrix $S_{(i,\sigma),(j,\sigma')}^{\alpha,\beta}$ as follows
\begin{equation}
\label{P_matrix}
P_{i,j}^{\alpha,\beta}(\epsilon,\vec{\theta})=\sum_{\sigma,\sigma'}\abs{S_{(i,\sigma),(j,\sigma')}^{\alpha,\beta}}^2 
\end{equation}
[see  Appendix~\ref{Scattering_matrix} for the details of the computation and Appendix~\ref{Symmetries} for the discussion of the special symmetries of the scattering coefficients in the symmetric case (i.~e. when $\Delta_L=\Delta_R$)]. The scattering coefficients $P_{i,j}^{\alpha,\beta}(\epsilon,\vec{\theta})$ represents the reflection ($i=j$) or transmission ($i\neq j$) probability of a QP of type $\beta$ injected from lead $j$ to end up as a QP of type $\alpha$ in lead $i$. We introduced the vector parameter $\vec{\theta}\equiv (\Phi,\phi)$ which includes both the magnetic flux $\Phi$ and the gauge invariant Josephson phase difference $\phi\equiv\phi_{\rm R}-\phi_{\rm L}$. We treat the two quantities independently, since $\Phi$ depends directly on the magnetic-field flux in the junction, while the phase bias  $\phi$ depends also on the dissipationless current imposed through the junction. They represent two different degrees of freedom needed to fully characterize the state of the TJJ.
It is worth to notice that
there is no dependence of the scattering coefficients on the contact potential~\cite{blonder_tinkham_1982} parameters $\Lambda_i$.~\footnote{This can be checked directly from the analytic expressions of the scattering coefficients, not reported here for brevity.
The independence of $\Lambda_i$ is a generic result valid also with other form -- not delta like -- of the interface potential at least in the absence of interaction.}. This is a direct consequence of the helicity of the edge channels which do not admit ordinary reflections at barriers (akin to the Klein paradox).~\cite{Lee19}

Finally, in Eq.~(\ref{current_Lambert_general}) we introduced also the function $q_i^{\alpha}(\epsilon)$ which represents the charge of the Bogoliubov QP of type $\alpha=\pm$ at energy $\epsilon$ at lead $i$ 
\begin{equation}
\label{qi}
q_i^{\alpha}(\epsilon)=\begin{cases} 
\alpha e & \text{for }\quad i=N  \\
[\Psi_{\alpha\mp}^{i,\chi_i}]^{\dagger} \mathcal{C}\Psi_{\alpha\mp}^{i,\chi_i} &  \text{for }\quad i=L/R ,
\end{cases}
\end{equation}
where $\mathcal{C}=e~\tau_3\otimes\sigma_0$ is the QP charge operator (being $e$ the electronic charge) expressed in terms of the Pauli matrices $\tau$ and $\sigma$ acting, respectively, on the particle-hole and spin space.
The index $\chi_i\equiv \chi_i(\epsilon,\epsilon_{DS})=\sign{(\epsilon-\abs{\Delta_i-\abs{\epsilon_{DS}}})}$, with $i=L,R$, refers to  
the different concavities of the eigenspectrum of Fig.~$\ref{dispersion}$, 
in agreement with the notation used in
Eq.~(\ref{Eigenfunctions}). 
More explicitly, using Eqs.~(\ref{Eigenfunctions}) and ($\ref{u_v}$), the second line of Eq.~(\ref{qi}) can be reduced to the usual formula 
\begin{equation}
q_{L/R}^{\pm}(\epsilon)=\pm e\left(\abs{u_{L\pm/R\mp}(\epsilon)}^2-\abs{v_{L\pm/R\mp}(\epsilon)}^2\right)
\end{equation}
representing the charge of Bogoliubov QPs, which reduces, for energy well above the gap, to $\pm e$ for an electron-like and a hole-like QP, respectively.

\section{Linear response regime}
\label{Linear_response_regime} 
The occurrence of a nonlocal thermoelectric response in the presence of a DS\cite{Blasi_2020_PRL} or of a phase difference\cite{Blasi_2020_Andreev_Inteferometer} was discussed and attributed to the helical nature of the edge states of the TJJ.
For simplicity, in those papers, only the symmetric case of equal gaps $\Delta_{\rm L}=\Delta_{\rm R}$ was considered.
We refer the reader to the Appendix~\ref{appendix_Symmetric_case} for a thorough discussion of such symmetric case, which complete the analysis done in Ref.~\onlinecite{Blasi_2020_Andreev_Inteferometer}
by presenting the analytical expressions of the heat and charge currents flowing through the probe obtained in the general case in which both the DS and the phase bias are present.

In this section we extend the discussion to the asymmetric case (where $\Delta_{\rm L}\ne \Delta_{\rm R}$).
This is not a mere generalization of previous results, since it provides important information relevant for realistic (experimental) realizations.
Indeed, when one of the two temperatures $T_i$ is bigger than $0.4 T_{C,i}$ (with $T_{C,i}$ being the critical temperature of the $i$-th superconductor) the self-consistent gap $\Delta_i(T_i)$ gets reduced from the zero-temperature value 
$\Delta_{i,0}$ and the right-left symmetric gap condition is hardly valid.~\cite{Tinkham}

Let us first clarify which is the relevant thermoelectrical response for the three terminal setup depicted in Fig.~\ref{scheme_with_temperature_voltage}.
We use the approach developed in Ref.~\onlinecite{Mazza14} to investigate thermoelectrical 
properties for multiterminal systems\cite{Benenti17}. The first step is to identify the independent currents.
Since charge and energy must be conserved, we are only left with 4 independent currents, out of 6 (3 charge currents and 3 heat currents).
In our three-terminal Josephson junction setup we consider the case of a stationary superconducting phase bias ($\dot{\phi}=0$), which indeed necessarily requires $V_L=V_R$.\footnote{This is an implicit consequence of the Josephson relation which states $\dot{\phi}=2e(V_L-V_R)/\hbar$.}
In such configuration the charge current flowing in the two superconductors is dominated by the dissipationless Josephson current which, in the linear response regime, is unaffected by the temperature difference between the electrodes.
In the following, in analogy to Refs.~\onlinecite{Blasi_2020_PRL,Blasi_2020_Andreev_Inteferometer}, we mainly discuss the only relevant currents which allow us to characterize the nonlocal thermoelectric response of our setup, which are the charge current ($J_N^-$) flowing in the normal probe and a heat current associated to the superconductors (see below for the proper definition).
Moreover, interesting features emerge also from the analysis of the linear heat current ($J_N^+$) flowing in the normal probe, which will be presented for completeness in Sec.~\ref{subsec:linear_heat_current_probe}.

\subsection{Nonlocal thermoelectric and Peltier coefficients}
\label{subsec:Nonlocal_thermoelectric_Peltier}

For small values of $V_N$ and $\delta T$, the charge current $J_N^-$ can be expanded up to the linear order in these quantities~\cite{Benenti17,Mazza14,RouraBas18,Hussein19,Rafa18,Kirsanov19}.
 The charge current for the probe reads 
\begin{equation}
\label{linear_JcN}
J_N^-=L_{11}^N \frac{V_N}{T} + L_{12}^N\frac{\delta T}{T^2},
\end{equation}
with $V_N/T$ and $\delta T/T^2$ taking the role of the relevant affinities of the problem, while $L_{11}^N$ and $L_{12}^N$ are linear transport coefficients.
We note that the electrical response to the bias $V_N$ is given by the conductance $G=L_{11}^N/T$, while the linear response to $\delta T$ (temperature 
difference between the superconductors) corresponds to a \emph{nonlocal} thermoelectric current, represented by the coefficient $L_{12}^N$.
It is important to stress now that in the linear-response regime the system is close to equilibrium.
As a result, the temperature of the probe needs to be set to its equilibrium value, namely $T_{N}=T$.
By setting $J_N^-=0$ in Eq.~(\ref{linear_JcN}) and solving for $V_N$ one finds  the linear Seebeck thermovoltage $V_N^{\rm S}$, through which we can compute the \emph{nonlocal} Seebeck coefficient $S=-V_N^{\rm S}/\delta T=(1/T) L_{12}^N/L_{11}^N$ similarly with the derivation of the local case.

Let us now consider the heat currents.
In the symmetric case  ($\Delta_{L}=\Delta_{R}$) and for $T_N=T$, the heat current flowing in left superconducting lead $J_L^+$ is exactly opposite to the current flowing in the right lead $J_R^+$, namely $J_L^+=-J_R^+$, while $J_N^+=0$ (i.e.~there is no heat current at the probe), due to the symmetries of the configuration, see Appendix \ref{appendix_Symmetric_case} for details.
When the left-right symmetry is broken ($\Delta_{L}\ne\Delta_{R}$), the two superconducting terminals are no more equivalent and the heat currents in the two superconducting terminals are different.
In such a case it is convenient to describe the heat current associated to the two superconductors by the average
\begin{equation}
\label{nonlocalheat}
J_{S}^+\equiv\frac{J_{L}^+-J_{R}^+}{2}.
\end{equation} 
So the linear-response regime is expressed by
\begin{equation}
\label{JSp}
J_{S}^+=L_{21}^S \frac{V_N}{T} + L_{22}^S\frac{\delta T}{T^2},
\end{equation}
The last term returns the \emph{local} heat conductance $\kappa=L_{22}^S/T^2$ between the two terminals, while the first term represents the \emph{nonlocal} Peltier-like contribution.
We notice that, in general, for a multiterminal system~\cite{Mazza14} one does not expect any specific symmetry relation between the nonlocal linear coefficients. However, with the definition of Eq.~(\ref{nonlocalheat}), the Onsager-Casimir relations~\cite{Onsager31,Casimir,Jacquod12,Benenti17} for the linear coefficients defined by Eqs.~(\ref{linear_JcN}) and (\ref{JSp}) can be expressed in the following form: $L_{12}^N(\vec{\theta})=-L_{21}^S(\vec{\theta})$.
In the next section we will numerically verify that such a relation holds independently of the ratio between the gaps. 

We will divide the analysis in two limiting situations. In the first case we will discuss the Onsager coefficients as functions of the DS in absence of any phase-bias, i.e. $\phi=0$. In the second case, instead, we analyze how in an extremely asymmetric case the thermoelectrical effect depends on the phase bias in absence of the DS ($\Phi=0$). In real experiments the discussed effects, which are different aspects of the nonlocal thermoelectricity, are  probably mixed but it is interesting to discuss them separately in order to clearly recognize their contributions to the nonlocal thermoelectrical signal.

\subsection{Asymmetric case for $\phi=0$}
\label{Linear response regime: asymmetric regime}
In this section we investigate the linear transport coefficients in the case where the right/left symmetry is broken by different zero-temperature superconducting gaps ($\Delta_{0,R}\ne\Delta_{0,L}$), and we define the ratio $r\equiv\Delta_{0,R}/\Delta_{0,L}=\xi_L/\xi_R$ where the second identity is expressed in terms of the coherence lengths $\xi_i=\hbar v_F/\pi\Delta_{0,i}$, with $i=R,L$.
We notice that, when $r\neq 1$, the scattering coefficients  $P_{i,j}^{\alpha,\beta}$ of Eq.~(\ref{P_matrix}) depend on the position of the probe.
For the sake of convenience, hereafter we consider  the probe positioned exactly in between the two superconductors (i.~e. $x_0=L/2$) and we fix $r\geq1$.~\footnote{The same results are obtained for $r\leq1$ by inverting the temperature bias and the sign of $\vec{\theta}$.}

\begin{figure}[h!!]
\centering
    \includegraphics[width=.49\textwidth]{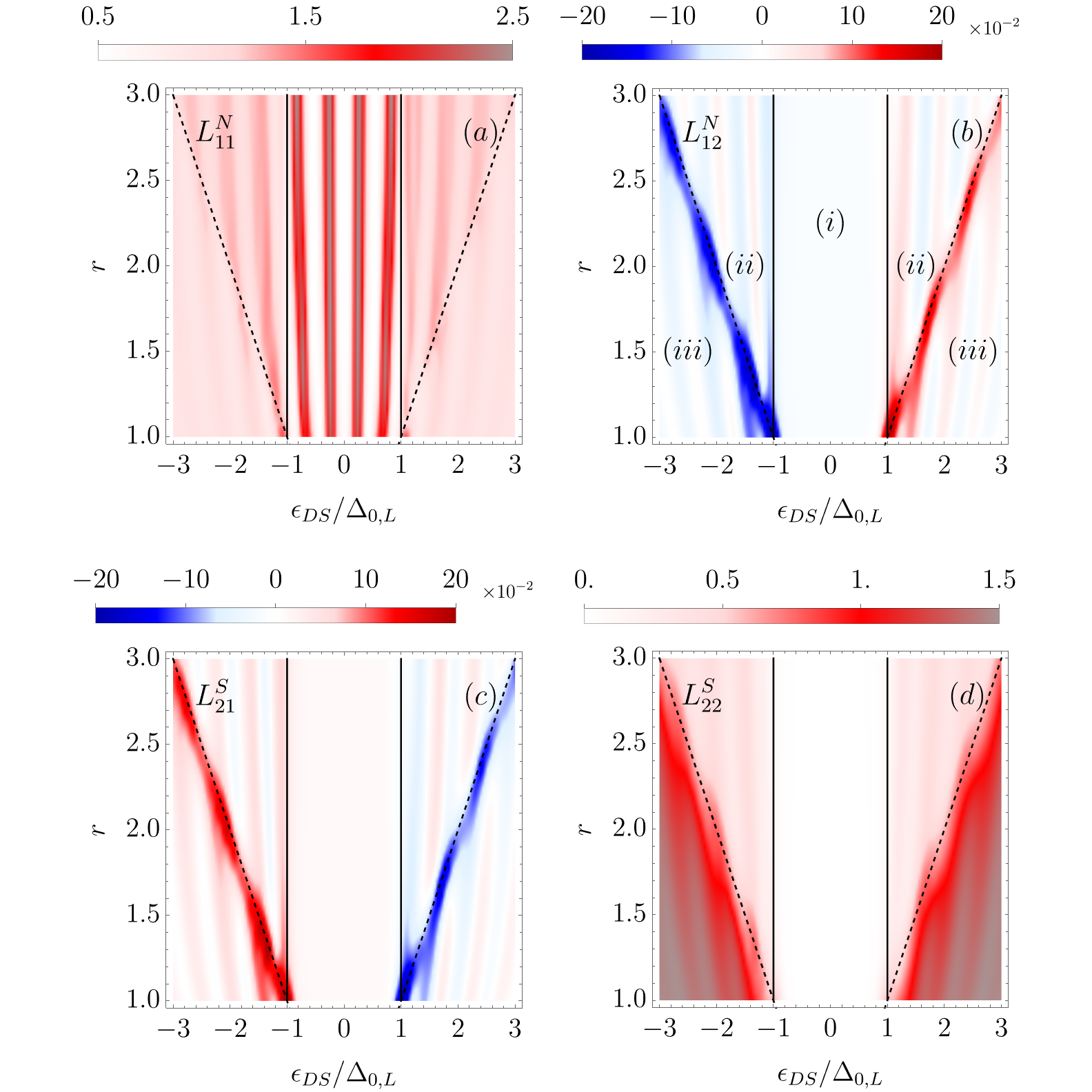}
    \caption{Onsager coefficients $L_{11}^N$ $(a)$, $L_{12}^N$ $(b)$, $L_{21}^S$ $(c)$ and $L_{22}^S$ $(d)$ as functions of $\epsilon_{DS}(\Phi)/\Delta_{0,L}$ and $r=\Delta_{0,R}/\Delta_{0,L}$ for $\phi=\phi_{\rm{R}}-\phi_{\rm{L}}=0$, $T/T_{C,L}=0.1$, $L/\xi_L =3$ and $\abs{t}^2=0.5$. Such quantities are normalized as follows: $L_{11}^N/(G_0 T)$, $L_{22}^S/(G_T T^2)$ and $(L_{12}^N, L_{21}^S)/(\sqrt{G_0 G_T T^3})$.} 
    \label{Onsager31_Different_Gaps} 
\end{figure}

The behavior of the linear coefficients $L_{ij}^{N/S}$ of Eqs.~(\ref{linear_JcN}) and (\ref{JSp}) is shown 
in Fig.~\ref{Onsager31_Different_Gaps} as a function of the DS $\epsilon_{DS}(\Phi)/\Delta_{0,L}$ and of the 
asymmetry parameter $r$, for fixed phase bias $\phi=0$. 
The local coefficients $L_{11}^N$ and $L_{22}^S$ are plotted in units of  $G_0T$ and $G_T T^2$, respectively, where $G_0$ denotes the electrical conductance quantum and $G_T=(\pi^2/3h)k_B^2 T$ the thermal one.
The nonlocal 
thermoelectrical coefficients $L_{12}^N$ and $L_{21}^S$ are plotted in units of $\sqrt{G_0 G_T T^3}$.
In these plots, we consider an intermediate coupling parameter to the probe ($\abs{t}^2=0.5$) and set the length of the 
junction three times longer than the superconducting coherence length, i.e. $L/\xi_L=3$.
Such a value of $L$ corresponds to the case of a long junction (given a coherence length $\xi_L$ in the proximized TI of the order of 600 nm), which allows the emergence of oscillations in the linear-response coefficients due to the proliferation of resonant states in the junction.
The choice of the value $\abs{t}^2=0.5$ comes from the fact that by increasing the coupling $\abs{t}^2$, the resonances get broadened, eventually disappearing when the coupling approaches unity, irrespective of the length $L$ (not shown).

Assuming $\Delta_{0,L}<\Delta_{0,R}$ (namely $r>1$), it turns out that the DS energy $\epsilon_{DS}(\Phi)$, modulated by the flux $\Phi$, modifies the dispersion curves of Fig.~\ref{dispersion} in a different way for the right and left superconductor.
In particular, for $\abs{\epsilon_{DS}(\Phi)}<\Delta_{0,L}$ the spectrum is gapped for energies
$\epsilon< \Delta_{0,L}-\abs{\epsilon_{DS}(\Phi)}$ in both superconducting leads. For $\Delta_{0,L}<\abs{\epsilon_{DS}(\Phi)}<\Delta_{0,R}$ we are in a situation where the gap is closed for the left superconductor, but open for the right one.
Finally when $\abs{\epsilon_{DS}(\Phi)}>\Delta_{0,R}$ the gaps are closed for both sides of the junction.

These different regimes can be recognized in the behavior of the Onsager coefficients, depicted in Fig.~\ref{Onsager31_Different_Gaps}, computed at low temperature $T/T_{C,L}=0.1$ (with $T_{C,L}$ the critical temperature of the left superconductor). In particular we recognize a subgap region (i) (between vertical solid lines) for $\abs{\epsilon_{DS}(\Phi)/\Delta_{L}}<1$, a partially gapped region (ii) (between the dashed and solid lines) and the supragap region (iii) (below diagonal dashed lines).
Notice that the clear distinction of the different regions emerging in the behavior of all the Onsager coefficients of Fig.~\ref{Onsager31_Different_Gaps} can be used, in principle, as an experimental observation of the different zero temperature gaps of the superconductors forming the junction.

For the electrical coefficient $L_{11}^N/G_0T$, depicted in Fig.~\ref{Onsager31_Different_Gaps}(a), we can recognize a similarity 
 with SINIS (superconductor-insulator-normal~metal-insulator-supercondutor) junctions where transport is typically suppressed in the subgap regime (i), though resonances are present (vertical red stripes) that
correspond to Andreev bound states (ABSs) crossing zero energy.
We checked that the linewidth of ABS 
resonances is dependent on the coupling parameter (by increasing the coupling $\abs{t}^2$ the resonances broaden). Furthermore, the number of resonances inside the region (i) grows with the ratio $L/\xi_L$ due to the proliferation of ABSs in the junction (not shown).
Similarly, in the partially gapped (ii) and the supragap (iii) regions we see a weak oscillating behavior (vertical stripes) reminiscent of Andreev interferometric effects.
On the other hand, we observe that the thermal coefficient $L_{22}^S/G_T T^2$ [Fig.~\ref{Onsager31_Different_Gaps}(d)] is completely suppressed in region (i) since, in this case, the ABS resonances cannot contribute to thermal transport due to the Andreev mirror effect.
Indeed, Andreev reflection does not allow the flow of energy through the S interface, since energy is carried by QPs, but not by the condensate.
However, the thermal conductance becomes finite and large in the supragap region (iii) where energy can be carried by QPs.
In the region (ii) the thermal transport between the two superconductors is not completely suppressed and it is influenced by an Andreev interferometric mechanism which determines an oscillating behavior.

Concerning the nonlocal coefficients $L_{12}^N$ and $L_{21}^S$, depicted in Fig.~\ref{Onsager31_Different_Gaps}(b) and (c) respectively, we confirm the validity of the discussed generalized Onsager symmetry which  
becomes $L_{12}^N(\Phi)=-L_{21}^S(\Phi)$ (being here $\phi=0$). The nonlocal coefficients clearly resemble some aspects of those discussed in Ref.~\onlinecite{Blasi_2020_PRL} for identical gaps. In particular, $L_{12}^N$ is suppressed in the subgap region (i) and two main peaks at $\abs{\epsilon_{DS}(\Phi)}=\Delta_{0,R}$ appear at the boundaries of regions (ii) and (iii) (dashed lines). 
In this condition, for $\epsilon_{DS}>0$ ($\epsilon_{DS}<0$), the top left (top right) band of the right proximized region in Fig.~\ref{dispersion}(c) nearly touch zero energy opening the possibility for thermoelectrical effects.
In particular, the helicity of the edge states allows (for example, for $\epsilon_{DS}>0$) a flow of hole-like QPs  [whose dispersion is represented by the dashed red curve in Fig.~\ref{dispersion}(a)] to move to the right from the proximized region S$_L$ and a flow of electron-like QPs [whose dispersion is represented by the solid blue curve in Fig.~\ref{dispersion}(c)] to move to the left from the proximized region S$_R$.
Under the application of a small temperature bias between the superconductors, the unbalance between the flow of cold electron-like QPs from the right and hot hole-like QPs from the left leads to a net thermoelectric current flowing through the probe~\cite{Blasi_2020_PRL}.

\begin{figure}[h!!]
\centering
    \includegraphics[width=.39\textwidth]{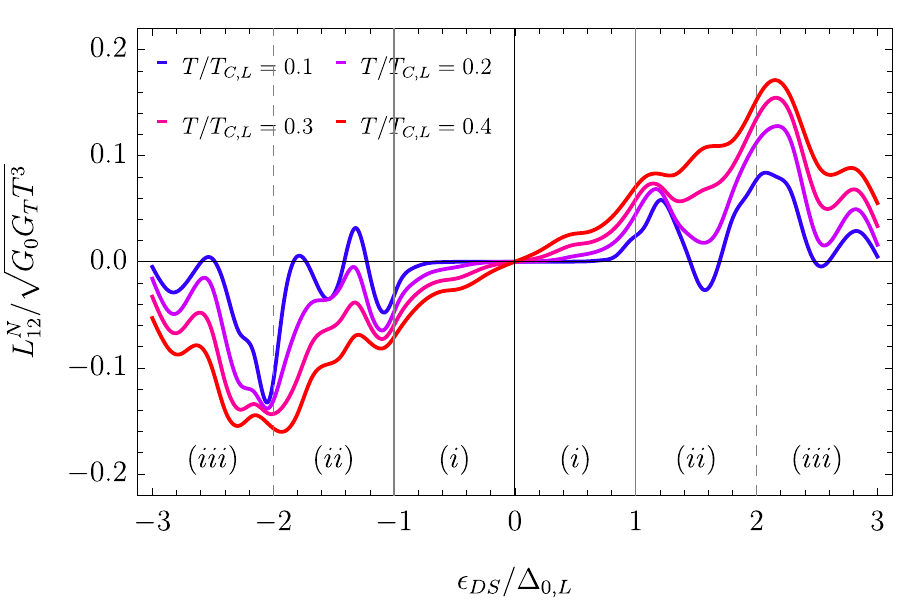}
    \caption{Nonlocal thermoelectrical coefficient $L_{12}^N$, expressed in units of $\sqrt{G_0 G_T T^3}$, as a function of $\epsilon_{DS}(\Phi)/\Delta_{0,L}$ for $r=2$ and different temperatures. We see that also in the subgap region $L_{12}^N\neq 0$ when temperature increase. Parameters are as in Fig.~\ref{Onsager31_Different_Gaps}.} 
    \label{L12_Different_Temp} 
\end{figure}

The true novel feature of the asymmetric case is represented by the appearance of a finite thermoelectric contribution ($L_{12}^N\neq 0$) in region (ii) [see Fig.~\ref{Onsager31_Different_Gaps}(b)], where the right superconductor is still  gapped. In such a case the situation resembles the Andreev interferometric mechanism, which appears at finite $\phi$ also with gapped bands.\cite{Blasi_2020_Andreev_Inteferometer} Indeed, even if in this case $\phi=0$, the flux $\Phi$ does not only generate a bands shifting due to DS but also affects the Andreev interference. This effect, in principle, is also present in the subgap region (i) even if it is exponentially suppressed as $\sim e^{-(\abs{\epsilon_{DS}(\Phi)}-\Delta_L)/k_BT}$. Indeed, only the supragap states contribute to thermoelectricity as long as they are thermally activated. This behavior is clarified by investigating the evolution of $L_{12}^N$ for different temperatures.
Figure~\ref{L12_Different_Temp} represents a horizontal cut of Fig.~\ref{Onsager31_Different_Gaps}(b) at $r=2$ for four different values of the temperature $T$ [the blue curve corresponds to the data plotted in Fig.~\ref{Onsager31_Different_Gaps}(b)].
When the temperature increases, $L_{12}^N$ also increases even in the subgap region (i). Furthermore, for higher temperatures all the resonances are smoothened due to the averaging between different energies.

At the lowest temperature $L_{12}^N$ presents changes of sign which disappear for higher values of $T$. 
This can be understood by noting that the sign of $L_{12}^N$ directly reflects the nature of carriers in the junction. 
For $k_BT\ll\Delta_{0,L}$ and for $\abs{\epsilon_{DS}}\lesssim \Delta_{0,R}$, only one type of carries, coming from the superconductor with the smaller gap, are allowed for the transport within the energy window $0\le \epsilon\lesssim k_B T$ (see Fig.~\ref{dispersion}): namely hole-like QPs (electron-like QPs) for $\epsilon_{DS}>0$ ($\epsilon_{DS}<0$). 
This flux of hole-like QPs (electron-like QPs) is contrasted by an opposite flux of electron-like QPs (hole-like QPs) originated due to Andreev reflections occurring at the interface with the other superconductor which is still gapped in that energy window. The (nontrivial) unbalance between these positive and negative charged carries determines the sign of the charge current flowing inside the probe and thus the sign of $L_{12}^N$.

As a final remark, it is important to notice that, differently from the results discussed in Ref.~\onlinecite{Blasi_2020_PRL} for identical gaps where $L_{12}^N$ was an odd function of $\Phi$ (namely $L_{12}^N(\Phi)=-L_{12}^N(-\Phi)$), in the asymmetric case (i.~e.~$\Delta_L\neq\Delta_R$) $L_{12}^N(\Phi)$ does not manifest any particular symmetry under the inversion $\Phi\rightarrow -\Phi$ (as clearly emerges from Fig.~\ref{L12_Different_Temp}). This is due to the fact that in the symmetric case there are extra symmetries in the scattering coefficients, as discussed in Appendix~\ref{Symmetries}.

\subsection{Asymmetric case for $\Phi=0$: interference effects on local thermoelectricity}
Let us first recall that in Ref.~\onlinecite{Blasi_2020_Andreev_Inteferometer} we have shown that a finite $\phi$ alone, i.e.~even in the absence of the DS, is sufficient to establish a nonlocal thermoelectric response due to an Andreev interferometric effect. 
Here we demonstrate that Andreev interferometry generates another peculiar effect that contributes to $L_{12}^N$, which emerges only by increasing the left/right asymmetry.
We consider the extreme situation in which $\Delta_{0,R}\rightarrow \infty$ (i.~e. $r\to \infty$), so that we can neglect all the contributions from the 
QPs on the right side, allowing us to obtain a simple analytical result for the charge current at probe $J_N^{-}$. 
In other words we are focusing on the regime where $k_BT_L,\Delta_L\ll\Delta_R$.
As a consequence of that choice, the temperature $T_R$ will not even enter in the discussion.
Therefore, the only thermal bias capable to drive a thermoelectric current through the probe is the 
local one present between the left superconductor S$_L$ and the probe N itself.
Interestingly, the right superconductor S$_R$, even if it does not directly contribute to the energy transport
(being fully gapped), still influences the exchange of charge between the N and S$_L$ through the Andreev reflection processes occurring at the interface with S$_R$.
The resulting dependence of the current $J_N^{-}$ on the phase difference $\phi$, see below, is due to the Josephson coupling established between the two superconducting leads.
Notice that this interference (coherent) effect on the local thermoelectricity between N and S$_L$ can be controlled by the application of a dissipationless current between the two superconductors, which changes the phase difference $\phi$.

The charge current at the probe takes the form
\begin{align}
\label{JNc_Mirror}
&J_N^-
&=\frac{2e}{h}\int_{0}^{\Delta_L}&\!\!\!d\epsilon~ F_{NN}^{+-}(\epsilon) \left[\mathcal{A}(\epsilon,\phi)+\mathcal{A}(\epsilon,-\phi)\right]~+\\
& &\frac{2e}{h}\int_{\Delta_L}^{\infty}\!\!\!d\epsilon~&
\!\!\!\!\sum_{\sigma,\sigma'=\pm}\sigma~ |r|^{|\sigma-\sigma'|}~F_{NL}^{\sigma,\sigma'}(\epsilon)~\mathcal{Q}(\epsilon,-\sigma'\phi)\nonumber ,
\end{align}
where the first term is the subgap energy contribution and the function
\begin{equation}
\label{Gamma0}
\mathcal{A}(\epsilon,\phi)=\frac{2\abs{t}^4\cdot \Theta(\Delta_L-\epsilon)}{1+\abs{r}^4+2\abs{r}^2\cos{\left(2\pi \frac{L\epsilon}{\hbar v_F}+\phi+\arcsin{\left(\frac{\epsilon}{\Delta_L}\right)}\right)}},\nonumber
\end{equation}
with $\abs{r}^2=1-\abs{t}^2$.
Such first term vanishes when $V_N=0$, since $F_{NN}^{+-}=0$ in that case (see Eq. (\ref{F_fermi_difference})).
The second term, instead, collects all the contribution for energies above the left gap and the function
\begin{equation}
\label{Qghost}
\mathcal{Q}(\epsilon,\phi)=\frac{\left(g(\epsilon)^2-1\right)\cdot \Theta(\epsilon-\Delta_L)}{g(\epsilon)^2+\abs{r}^4-2g(\epsilon)\abs{r}^2\sin{(2\pi \frac{L\epsilon}{\hbar v_F}+\phi)}} ,
\end{equation}
with $g(\epsilon)=e^{\arccosh(\epsilon/\Delta)}$.
Note that in Eq.~(\ref{JNc_Mirror}) the function $F^{\alpha\beta}_{NL}=f_N^\alpha(\epsilon)-f_{L}^\beta(\epsilon)$ involves only the electrodes N and S$_L$.
From Eq.~(\ref{JNc_Mirror}) it is possible to derive the analytical expressions of the Onsager coefficients $L_{11}^N$ and $L_{12}^N$ of Eq.~(\ref{linear_JcN}) in the case $V_N,\delta T\to0$:
\begin{align}
\label{L11}
L_{11}^N
&=-\frac{2e^2}{h}\int_{0}^{\Delta_{L}}\!\!\!\!\!\!d\epsilon~2T f_0'(\epsilon) \left[\mathcal{A}(\epsilon,\phi)+\mathcal{A}(\epsilon,-\phi)\right]~+\\
&-\frac{2e^2}{h}\int_{\Delta_{L}}^{\infty}\!\!\!\!\!\!d\epsilon~T f_0'(\epsilon)(1+\abs{r}^2)\left[\mathcal{Q}(\epsilon,\phi)+\mathcal{Q}(\epsilon,-\phi)\right]\nonumber
\end{align}
\begin{equation}
\label{L12Ghost}
L_{12}^N=-\frac{2e}{h}\int_{\Delta_{L}}^{\infty} d\epsilon~\epsilon~T f_0'(\epsilon)\abs{t}^2\left[\mathcal{Q}(\epsilon,\phi)-\mathcal{Q}(\epsilon,-\phi)\right]
\end{equation}
where $f_0'(\epsilon)=\left[4k_BT \cosh{\left(\epsilon/2k_B T\right)}^2\right]^{-1}$, is the derivative of the Fermi function $f_0(\epsilon)=\left[1+e^{\epsilon/k_BT}\right]^{-1}$.
From Eqs.~($\ref{L11}$) and ($\ref{L12Ghost}$) clearly emerges the even and odd parity in $\phi$ of $L_{11}^N$ and $L_{12}^N$, respectively.
\begin{figure}[ht]
\centering
    \includegraphics[width=0.49\textwidth]{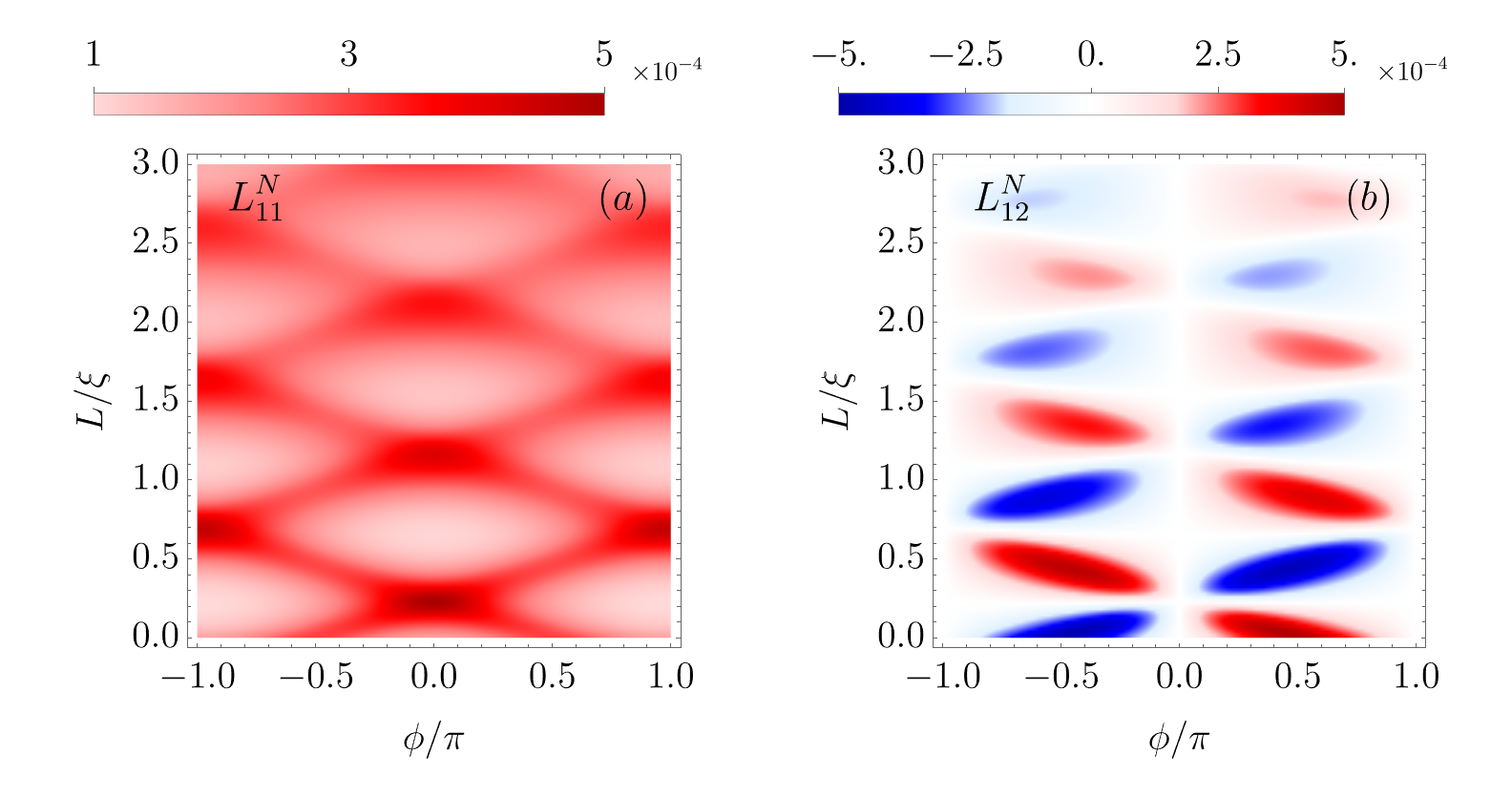}
    \caption{Charge current $J_N^-$ linear coefficients $L_{11}^N$ and $L_{12}^N$ in the extreme asymmetric limit $r\to\infty$ as a function of the phase bias $\phi$ and junction length $L$. Other parameters are $T/T_{C,L}=0.2$, $|t|^2=0.5$} 
    \label{L11_L22_DiffGaps} 
\end{figure}
In Fig.~\ref{L11_L22_DiffGaps} we plot those two quantities with respect to the phase bias $\phi$ and the junction length $L$.
The main difference with respect to the purely nonlocal thermoelectric coefficient of Ref. \onlinecite{Blasi_2020_Andreev_Inteferometer} is that the thermoelectric coefficient $L_{12}^N$ of Eq. (\ref{L12Ghost}) depends on $\sin{(2\pi \frac{L\epsilon}{\hbar v_F}+\phi)}$ (see Eq. (\ref{Qghost})) instead of $\cos{(2\pi \frac{L\epsilon}{\hbar v_F}+\phi)}$ (see Appendix \ref{appendix_Symmetric_case}).
For long junctions (i.~e.~$L\to\infty$) the contributions for different energies average to zero recovering the expected result of a standard $NS$ junction. This clearly shows that when the superconducting leads are far from each other, the Andreev interferometric mechanism on the local thermoelectrical transport is suppressed.

We conclude this section by observing that the \emph{local} thermoelectric effect in the probe, which emerges with gap asymmetry, represents one of the main source of disturbance to the measurement of the \emph{nonlocal} thermoelectric effect discussed in the sections above. This implies that, in order to clearly see the nonlocal thermoelectrical effect, it is convenient to be in a situation with weak asymmetry. However, we estimated that this local thermoelectric effect is of the order of few $\mu$V/K, to be compared with the nonlocal thermoelectric effect determined by DS, discussed before, which could reach values of the order of many tenths $\mu$V/K (see below).
So we do not expect that spurious local thermoelectrical effects would substantially affect the nonlocal measurements at least for moderate gap asymmetry.

\subsection{Linear-response heat current at the probe}
\label{subsec:linear_heat_current_probe}
\begin{figure}[h!!]
\centering
    \includegraphics[width=.49\textwidth]{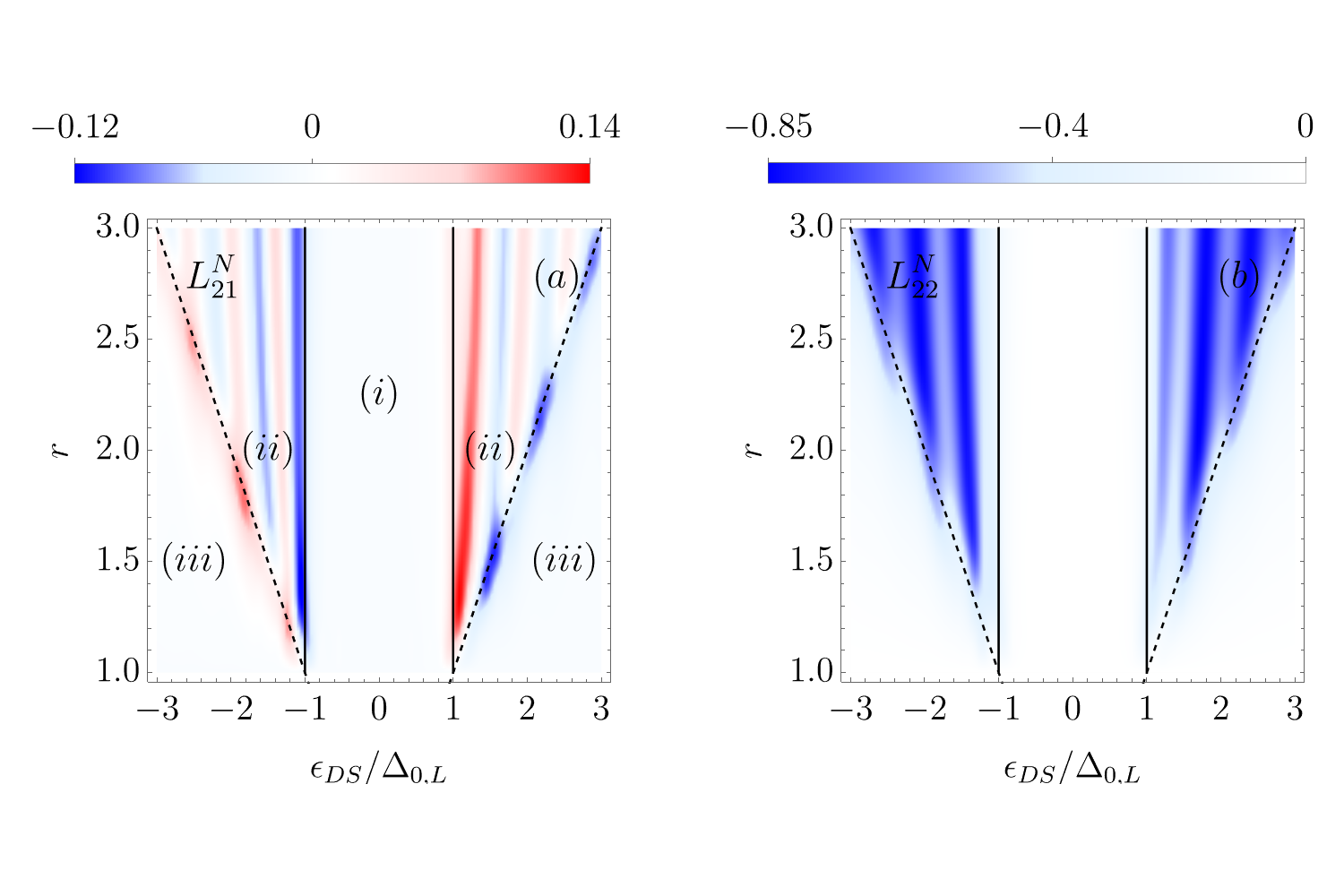}
    \caption{Heat current $J_N^+$ Onsager coefficients $L_{21}^N$ $(a)$, $L_{22}^N$ $(b)$ as functions of $\epsilon_{DS}(\Phi)/\Delta_{0,L}$ and $r=\Delta_{0,R}/\Delta_{0,L}$, for $\phi=\phi_{\rm{R}}-\phi_{\rm{L}}=0$, $T/T_{C,L}=0.1$, $L/\xi_L =3$ and $\abs{t}^2=0.5$. Such quantities are normalized as follows: $L_{21}^N/(\sqrt{G_0 G_T T^3})$, $L_{22}^S/(G_T T^2)$.} 
    \label{Fig_linear_JNh} 
\end{figure}

In the asymmetric case ($\Delta_{R}\ne\Delta_{L}$) it is interesting to discuss the behavior of the heat current flowing in the probe $J^+_N$.
Similarly to Eq.~(\ref{linear_JcN}), in the linear-response regime we can write
\begin{equation}
\label{linear_JhN}
J_N^+=L_{21}^N \frac{V_N}{T} + L_{22}^N\frac{\delta T}{T^2},
\end{equation}
The coefficient $L_{21}^N$ accounts for the local Peltier effect at the probe (describing how the heat current at the probe is influenced by the voltage bias $V_N$). While $L_{22}^N$ represents the transverse heat response at the probe, i.e. the heat current induced in the probe as determined by a trasversal temperature gradient \emph{between} the superconductors.
In contrast to the symmetric case in which $J_N^+=0$ in the linear response regime (see Appendix~\ref{appendix_Symmetric_case} and Refs.~\onlinecite{Blasi_2020_Andreev_Inteferometer,Blasi_2020_PRL}), in the asymmetric configuration $J_N^+$ can be finite even in linear response. This can be interpreted as the consequence of the contributions of the different heat currents deriving from the two superconducting leads kept at different temperatures. In Fig.~\ref{Fig_linear_JNh} we plot $L_{21}^N$, panel (a), and $L_{22}^N$, panel (b), as functions of $\epsilon_{DS}$ and $r$; both quantities are significantly different from zero only in region $(ii)$. 
Indeed in region (i), $J_N^+$ vanishes since the system it is not able to activate enough supragap states.\footnote{Note that the Joule component can be neglected in the linear regime since it scales at least as $V_N^2$}
On the other hand, in region (iii) the two gaps are both closed and there is almost no difference in the thermal coupling between right and left leads. 
Since the probe is at the average temperature $T=(T_R+T_L)/2$, the system behaves very similarly to the symmetric case ($r=1$ bottom border of the figures) where the thermal flux from hotter lead is compensated with the thermal losses in the colder lead giving a null net thermal current in the probe.

Furthermore, we notice the behavior of the transverse thermal coefficient $L_{22}^N$ of the system can be described in terms of a thermal divider, i.e.~a device that controls the sign of the heat current in an intermediate third terminal (the probe) assuming that there is a main heat flow generated from the thermal gradient $\delta T$ (between the superconductors).\footnote{The name is inspired by an evident analogy with voltage dividers}
In particular the three terminal setup can be described as a series of two thermal conductances with the probe in the middle. In such a case, the heat flux in the probe would depend essentially on the ratio of thermal resistances.
Indeed, in region (ii) for $r>1$ the thermal coupling of the probe with right lead is very opaque with respect to the left lead. Since the probe temperature is fixed at $T=(T_R+T_L)/2$ there is more heat flowing from the left lead than into the right one, determining the negative sign of $L_{22}^N$ (since positive thermal current of the probe is defined exiting from the probe).\footnote{The negative sign of this quantity is simply determined by the adopted convention for probe currents but clearly the spontaneous thermal flux always flows from the hotter to the colder lead, in agreement with thermodynamic laws.}   We verified that the sign of $L_{22}^N$ changes globally for the opposite case $r<1$ (not shown). This behavior qualitatively resembles the concept of the thermal router\cite{Timossi18} in superconducting hybrid systems. 

In region (ii), we notice that both linear coefficients are also oscillating (see the vertical stripes). In particular, the sign changing of $L_{21}^N$ with $\Phi$ reflects the change of the main carrier (electron-like or hole-like QP) as determined by Andreev interference discussed in Sec.~\ref{Linear response regime: asymmetric regime}. It is indeed important to remind that $L_{21}^N$ being a local Peltier-like coefficient its sign will be directly dependent to the sign of the dominant carrier.
Instead $L_{22}^N$ has not similar sign changes being associated to a transversal thermal response that cannot distinguish on the main carrier charge sign.

\section{Non-linear response regime}
\label{Non-linear response regime}

\begin{figure*}[ht]
\centering
    \includegraphics[width=1\textwidth]{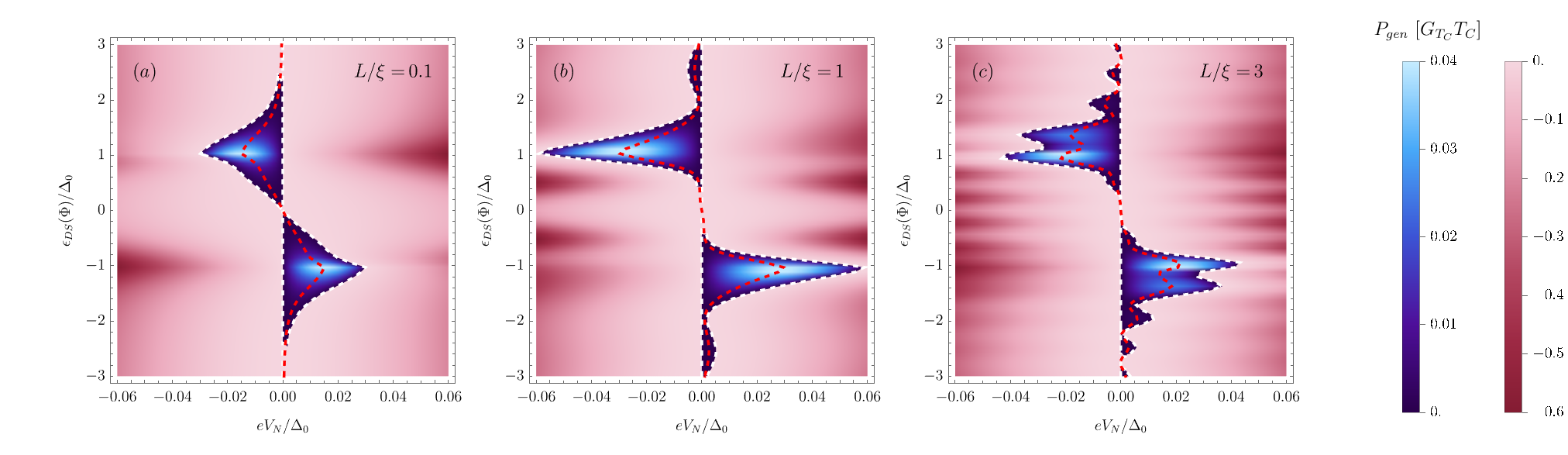}
    \caption{Electrical power $P$ in units of $G_{T_C}T_C$ as a function of $eV_N/\Delta_0$ versus $\epsilon_{DS}(\Phi)/\Delta_0$ in the case of short $(a)$ $L/\xi=0.1$, medium $(b)$ $L/\xi=1$ and long $(c)$ $L/\xi=3$ junction. In blue is depicted the generated power $P_{gen}=P>0$. The white dashed line corresponds to the stopping voltage curve, while the red dashed line indicates the maximum generated power $P_{max}$. Here we chosen $\abs{t}^2=0.5$ (an intermediate coupling parameter representing a not fully Ohmic contact with the probe), $\delta T/T_C\approx 0.4$ (large enough to guarantee the highest possible electrical power by keeping constant and equal the gaps of the superconductors) with $T/T_C=0.2$ and $\phi=0$.} 
    \label{Power_L} 
\end{figure*}
In this section we investigate the behavior of the nonlocal thermoelectricity from the perspective of a  thermodynamic engine within the nonlinear regime.  
The laws of thermodynamics set very general constraints on the currents of Eq.~(\ref{current_Lambert_general}).
In particular, the first law of thermodynamics, which guarantees energy conservation, can be written as
\begin{equation}
\sum_i{J_i^+}=P,
\end{equation}
where $P$ is the electrical power 
\begin{equation}
\label{P}
P=-\sum_i{J_i^- V_i}.
\end{equation}
With this definition, $P$ is positive when the current flows against the applied bias, i.~e. there is a thermopower generated in the system that can be dissipated on an external load ($P\equiv P_{gen}$).
The device thus works as a thermoelectrical engine.\cite{Whitney13,Whitney14,Sothmann14,Mazza14,Benenti17}
Notice that, since we set $V_{L}=V_{R}=0$, Eq.~(\ref{P}) reduces only to $P=-J_N^-V_N$, which means that the power is dissipated in the probe circuit only.
Another important performance quantifier is the efficiency defined as
\begin{equation}
\label{efficiency}
\eta
=\frac{P_{gen}}{\sum_i^+{J_i^+}},
\end{equation}
where the numerator corresponds to the electrical power generated $P_{gen}$, while the denominator corresponds to the total heat current entering the system (the apex $+$ in the sum means that we are summing only positive heat currents).
In the remaining part of the paper we discuss the symmetric case ($r=1$), with $\Delta_0\equiv \Delta_{0,L}=\Delta_{0,R}$ and $T_C\equiv T_{C,L}=T_{C,R}$.
However, when the temperature difference between the two superconductors becomes comparable with the critical temperatures, the  superconducting gaps on the two sides of the junction will take different values. 
In what follows we take into account this fact by including self-consistent temperature dependence of the two gaps.

\subsection{Electrical Power and maximum Power}
\label{Electrical_Power_and_maximum_Power}
The electrical power $P$ of Eq.~(\ref{P}) is presented in Fig.~$\ref{Power_L}$ and expressed in units of $G_{T_C}T_C$ (with $G_{T_C}=(\pi^2/3h)k_B^2T_C$ the thermal conductance quantum at $T_C$), as a function of $eV_N/\Delta_0$ and $\epsilon_{DS}(\Phi)/\Delta_0$ in the case of short $L/\xi=0.1$ (Fig.~$\ref{Power_L}$~$(a)$), medium $L/\xi=1$ (Fig.~$\ref{Power_L}$~$(b)$) and long $L/\xi=3$ (Fig.~$\ref{Power_L}$~ $(c)$) junctions, with $\xi=\hbar \varv_F /\pi \Delta_0$.
Here we set $\abs{t}^2=0.5$, representing a intermediate Ohmic contact with the probe, and a phase difference $\phi=0$. 
It is important to notice that here we set $T=0.2T_C$ and $\delta T/T_C\approx 0.4$, which is the largest thermal bias for which $T_L,T_R\lesssim 0.4 T_C$, such that the superconducting gaps can be safely considered still constant [$\Delta_{L}(T_L)\approx \Delta_{R}(T_R)\approx \Delta_0$].

In Fig.~$\ref{Power_L}$ the white dashed lines represent the stopping voltage $V_{stop}$ defined through the equality $P_{gen}(V_{stop})=J_N^-(V_{stop})=0$, while the red dashed lines locate the maximum generated power $P_{max}=\max_{V_N}[P_{gen}]$.
By comparing Figs.~$\ref{Power_L}$~$(a)$, $(b)$ and $(c)$, we can see that the behavior of the electrical power for different lengths of the junction remains roughly the same. In particular, when the gap closes due to the flux bias $\Phi$, i.e. when $\abs{\epsilon_{DS}(\Phi)/\Delta_0}\approx 1$, the electrical generated power $P_{gen}$ is maximized irrespective of the length $L$.
For long junctions, the only additional feature is the presence of ripples in the generated power due to the proliferation of resonant states inside the junction [see Fig.~$\ref{Power_L}$~$(c)$] which in turn affects also the supragap states.
No oscillations occur at any lengths when $|t|^2\approx 1$ (not shown).
Furthermore, another important feature which emerges from Fig.~$\ref{Power_L}$, is that the sign of the stopping voltage $V_{stop}$  changes when $\Phi\rightarrow -\Phi$ as a consequence of the antisymmetry of the thermoelectricity under the magnetic field inversion.

\begin{figure}[h!!]
\centering
    \includegraphics[width=.4\textwidth]{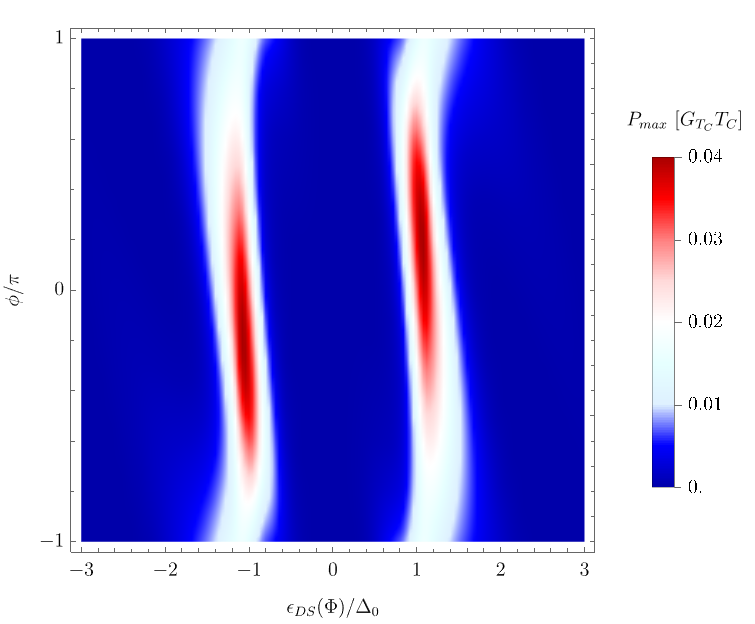}
    \caption{Maximum generated power $P_{max}$ in units of $G_{T_C}T_C$ as a function of $\epsilon_{DS}(\Phi)/\Delta_0$ versus $\phi/\pi$. Here we considered $L/\xi=1$, $\abs{t}^2=0.5$ and $\delta T/T_C\approx 0.4$ (large enough to guarantee the highest possible electrical power by keeping constant and equal the gaps of the superconductors) with $T/T_C=0.2$.} 
    \label{Pmax_p_phi} 
\end{figure}

A study of the dependence of the maximum generated power
on the phase difference $\phi$ is presented in Fig.~$\ref{Pmax_p_phi}$.
Here, $P_{max}$ is plotted in units of $G_{T_C}T_C$, as a function of the two external tuneable knobs, i.~e.~$\epsilon_{DS}(\Phi)/\Delta_0$ and $\phi/\pi$, setting $L=\xi$. This length is realistic assuming a scanning tunneling microscopy (STM) tip with a state-of-the-art width of 100 nm and a coherence length $\xi$ in the proximized TI of the order of 600 nm.~\cite{Bocquillon17}
\begin{figure}[h]
\centering
    \includegraphics[width=.49\textwidth]{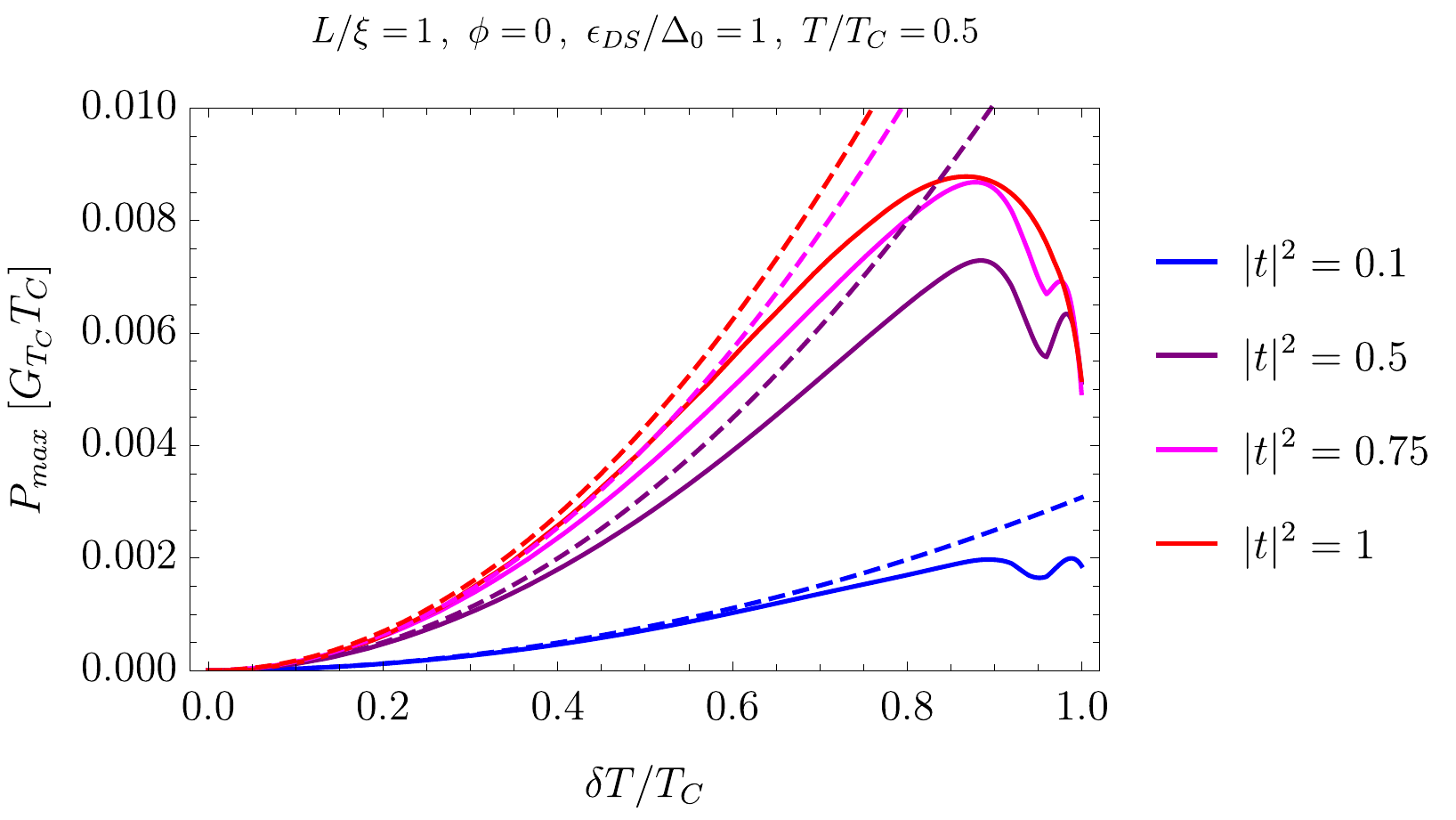}
    \caption{Maximum generated power $P_{max}$ in units of $G_{T_C}T_C$ as a function of $\delta T/T_C$ for different values of the coupling parameter $\abs{t}^2$ (solid lines) in comparison with the maximum power in the linear regime (dashed lines). Other parameters in the heading. Here we chosen $T/T_C=0.5$ in order to maximize the range of the thermal gradient $\delta T/T_C$ so that the gap of the hotter superconductor does not close (i.~e.~ $T_{L}/T_C\lesssim 1$). } 
    \label{PmaxSYM} 
\end{figure}
Furthermore, over a length $L\sim\xi$ no backscattering events are expected to occur at the operating temperatures for our setup, typically of a few kelvin. Also in this case we consider an intermediate coupling parameter $\abs{t}^2=0.5$ with the probe and a thermal gradient $\delta T/T_C\approx 0.4$ with $T/T_C=0.2$. 
Importantly, Fig.~\ref{Pmax_p_phi} shows that the following symmetry holds: $P_{max}(\vec{\theta})=P_{max}(-\vec{\theta})$. The same holds, in general for the electrical power, i.~e.~$P(\vec{\theta},V_N)=P(-\vec{\theta},-V_N)$.

In Fig.~$\ref{PmaxSYM}$ we present the results of the maximum generated power $P_{max}$ as a function of $\delta T/T_C$ for different values of the coupling parameter $\abs{t}^2$. 
In this case,  we explicitly consider the temperature dependence of the superconducting gaps by using the approximated formula $\Delta_i=\Delta_{0}\tanh{\left(1.74\sqrt{T_{C}/T_i-1}\right)}$ (with $i=L,R$) which is accurate better than $2\%$ with respect to the self-consistent BCS result.~\cite{Kamp19,Tinkham}
We compare it with the maximum power in the linear response regime given by the relation $P_{max}=\frac{GS^2}{4}\delta T^2=\frac{L_{12}^2}{L_{11}}\frac{\delta T^2}{4 T^3}$(dashed lines).~\cite{Benenti17}
In the figure we consider $T/T_C=0.5$ in order to maximize the excursion of the thermal gradient $\delta T/T_C\in[0,1]$ by preserving the superconducting state of the leads (namely, such that the gap of the hotter superconductor does not close, i.~e.~$T_{L}/T_C\lesssim 1$). Here we set $L/\xi=1$, $\phi=0$ and $\epsilon_{DS}(\Phi)/\Delta_0=1$. 
From an analysis of the result of Fig.~$\ref{PmaxSYM}$ emerges that all the curves match the trend of the linear regime for small $\delta T$ as expected. Nonlinearities emerge only for $\delta T/T_C\gtrsim 0.4$ due to the closure of the gap of the hotter superconductor.
\begin{figure}
  \centering
		\subfigure[][]{\label{efficiency_SYM}\includegraphics[width=1\columnwidth]{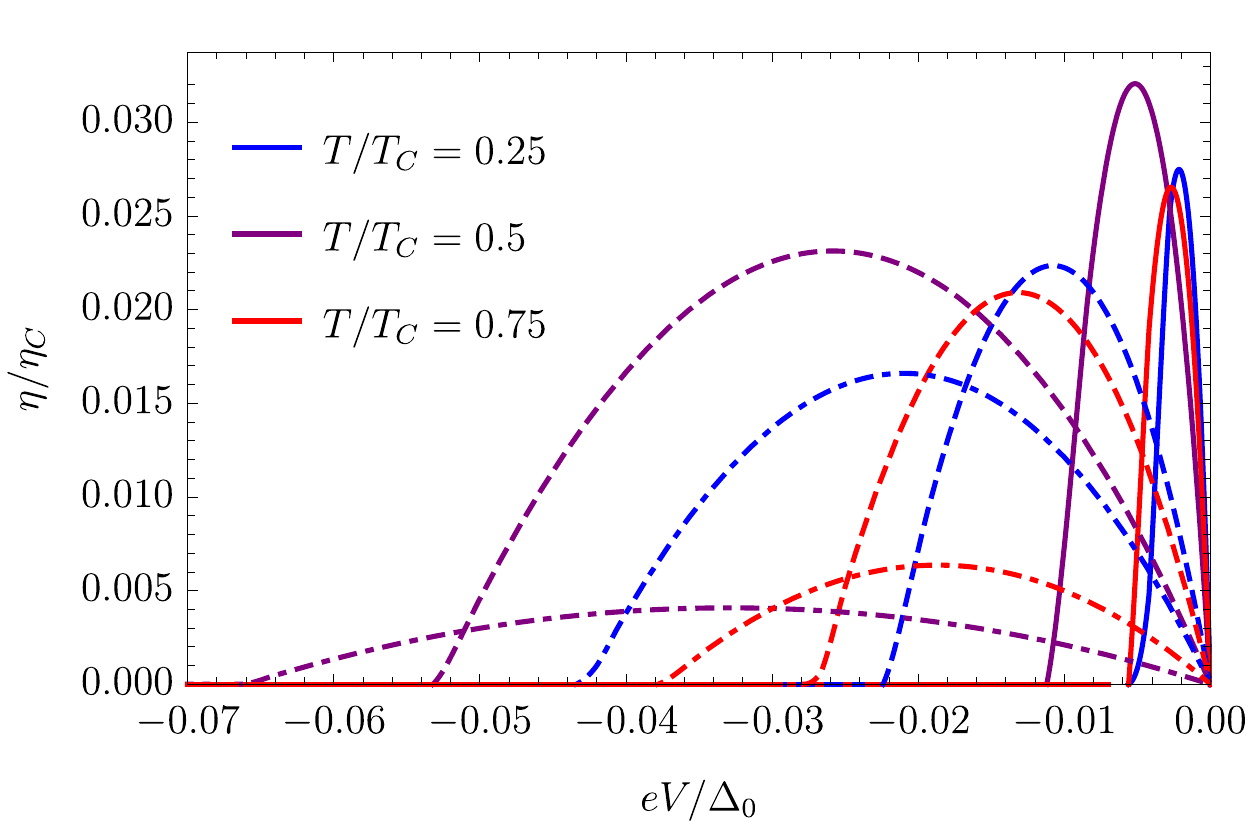}}
		 \hfill
		\subfigure[][]{\label{Lasso_SYM}\includegraphics[width=1\columnwidth]{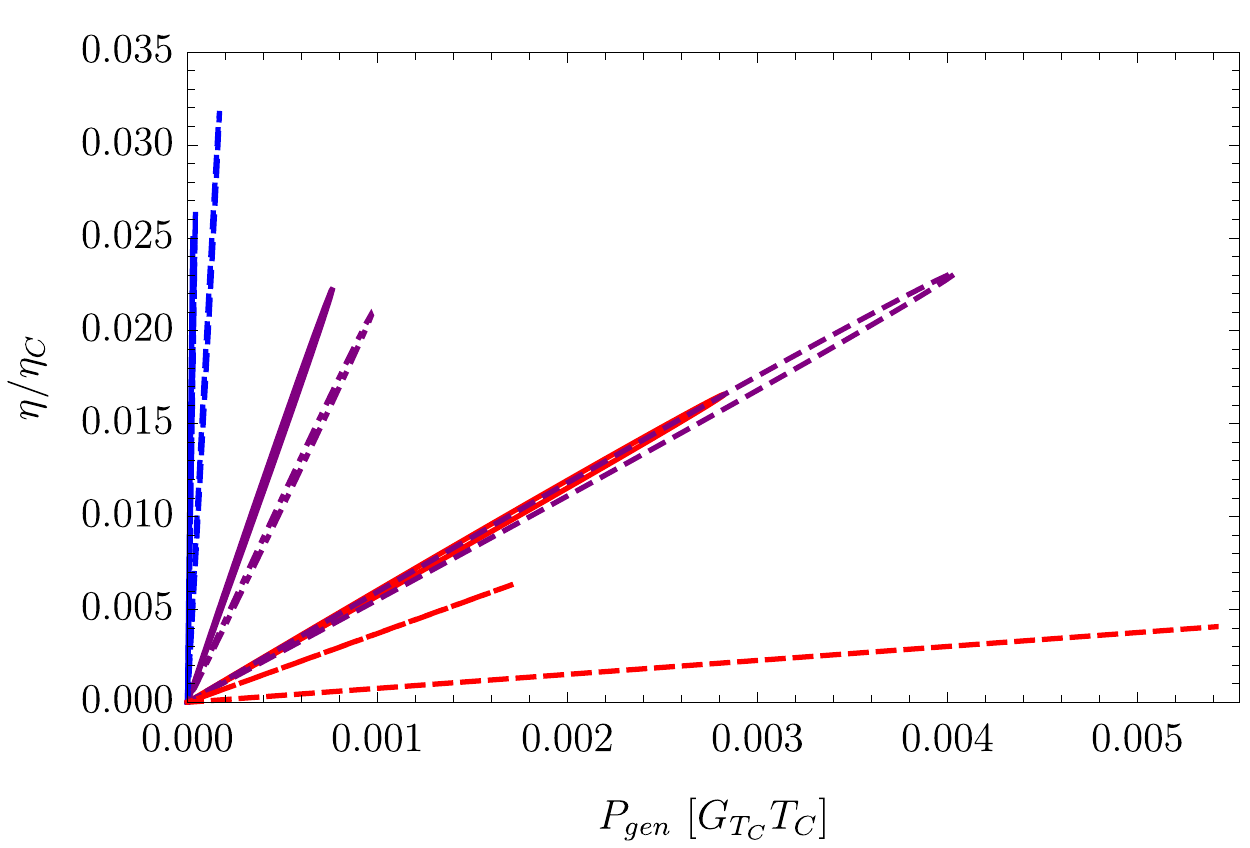}}
  \caption{(a) - Efficiency $\eta$ normalized with respect the Carnot efficiency $\eta_C=\delta T/T$ as a function of $eV_N/\Delta_0$. (b) - So-called lasso diagrams showing the normalized efficiency $\eta/\eta_C$ at every power output expressed in units of $G_{T_C} T_C$. Different colors correspond to different probe's temperature $T/T_C=0.25$ (blue line), $T/T_C=0.5$ (yellow line), $T/T_C=0.75$ (red line). Different style of lines correspond to different values of the thermal gradient (normalized with respect to $\delta T_{max}$ of Eq.~(\ref{dT_max}): $\delta T/\delta T_{max}=0.1$ (solid line), $\delta T/\delta T_{max}=0.5$ (dashed line), $\delta T/\delta T_{max}=1$ (dot-dashed line).}
  \label{eta_Lasso}
\end{figure}

It is important to notice that the red curves of Fig.~$\ref{PmaxSYM}$, corresponding to the case of perfect coupling with the probe ($|t|^2= 1$), do not depend on neither the phase difference $\phi$ nor the length of the junction $L$.
This curve clearly maximizes the performance with respect to all the other cases $|t|^2<1$.

\subsection{Efficiency and Lasso diagram}
\label{Efficiency_and_Lasso_diagram}
We now present in Fig.~$\ref{eta_Lasso}$, the results for the efficiency, defined in Eq.~(\ref{efficiency}), by assuming perfect coupling with the probe $\abs{t}^2=1$ and $\epsilon_{DS}(\Phi)/\Delta_0=1$, for which we have maximum power (according to the discussion of the previous section).
In this situation, both the efficiency $\eta$ and $P_{max}$ do not depend on the phase difference $\phi$ and the junction length $L$. Here, we consider different values of the temperature of the probe, namely $T=0.25~T_C,~0.5~T_C$ and $0.75~T_C$, corresponding to different colors in Fig.~$\ref{eta_Lasso}$. For each value of $T$, we take three different values of the thermal bias $\delta T=0.1~\delta T_{max},0.5~\delta T_{max},\delta T_{max}$ (corresponding to the different style of the lines: solid, dashed, dotdashed).
$\delta T_{max}$ take the following value
\begin{equation}
\label{dT_max} 
\delta T_{max} = \min{\{2 T,2 (T_C-T)\}} 
\end{equation}
in order to always fulfill the condition $0<T_L,T_R<T_C$ for any operating temperature of the superconductors.
All the curves in Fig.~$\ref{efficiency_SYM}$ present the same ``reversed-parabola" behavior, passing through $V_N=0$  when the themocurrent becomes null again, similarly to what is expected for a linear thermoelectrical engine\cite{Benenti17}. The nonlocal thermoelectrical engine has a quite small maximum value for the efficiency, i.~e. $\eta/\eta_C\lesssim 3.5\%$. 
This low efficiency can be attributed to the large flux of heat entering the system [thus increasing the denominator of Eq.~(\ref{efficiency})] which occurs when gaps close as a consequence of the DS.

A convenient way to present the efficiency at a given power output, and vice versa, is in the form of lasso diagrams as depicted in Fig.~$\ref{Lasso_SYM}$. The parameter that is changed along the lasso-line is the applied voltage $V_N$. As we can notice, the lasso-curves in Fig.~$\ref{Lasso_SYM}$ are long and narrow loops for all values of the controlling parameters ($T$ and $\delta T$). This implies that maximum efficiency and maximum output power occur at the same value of parameters. This is advantageous for the operation of a thermoelectric device, where one typically must decide whether to optimize the engine operation with respect to efficiency or power output and constitute a major difference between this nonlocal thermoelectrical engine and standard linear thermoelectrical engines.~\cite{Benenti17}

\subsection{Nonlinear Seebeck coefficient}
\label{Nonlinear_Seebeck}

\begin{figure}[h]
\centering
    \includegraphics[width=.49\textwidth]{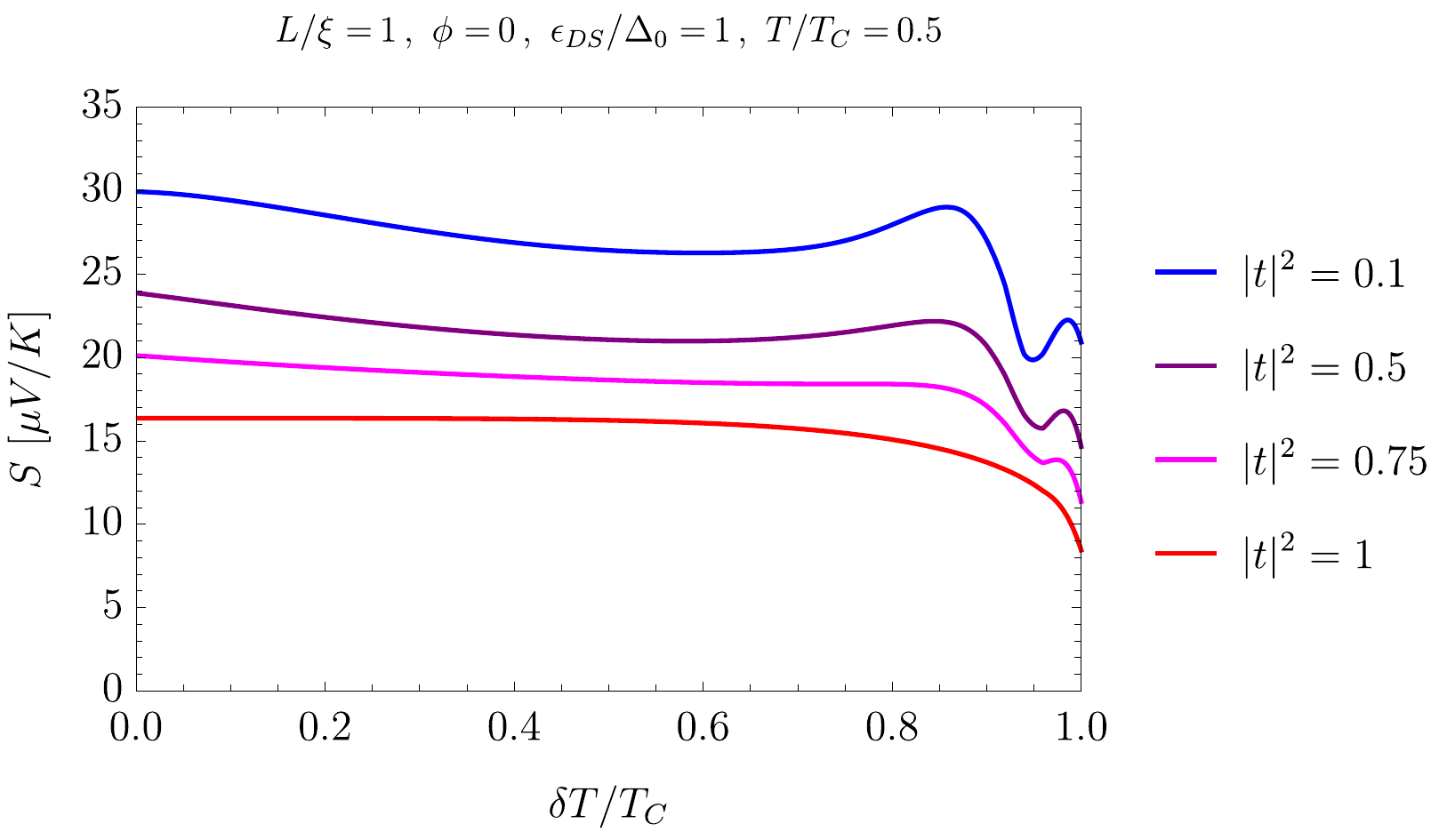}
    \caption{Seebeck coefficient in units of $\mu V/K$ as a function of $\delta T/T_C$ for different values of the coupling parameter $\abs{t}^2$. The other parameters are the same of Fig.~$\ref{PmaxSYM}$ (see the heading).} 
    \label{Seebeck_L_1} 
\end{figure}

In this section we present the result of the nonlinear Seebeck coefficient as a function of $\delta T/T_C$ for different values of the coupling parameter $\abs{t}^2$.
In Fig.~$\ref{Seebeck_L_1}$ we plot the nonlinear nonlocal Seebeck coefficient defined as
\begin{equation}
S=-\frac{V_{stop}}{\delta T}
\label{Seebeck}
\end{equation}
expressed in units of $\mu V/K$, where $V_{stop}$ is the stopping voltage for which $J_N^-(V_{stop})=P_{gen}(V_{stop})=0$ (see white dashed lines of Fig.~$\ref{Power_L}$).
We observe that the nonlinear Seebeck coefficient is quite big, considering that the operating temperature for these devices is of the order of few kelvin. Moreover, we see that $S$ weakly depends on the temperature difference as long as the gap remains unaffected. Furthermore, we see that $S$ increases going toward the tunnelling limit $|t|^2 \to0$. We can conclude that nonlocal thermoelectricity is a strong effect.

\section{Conclusions}
\label{Conclusions}
We have discussed the nonlocal thermoelectricity generated in a three terminal topological Josephson engine where one edge of a 2D TI is coupled to a normal metal probe. The nonlocal thermoelectricity is associated to the helical nature of the edge states and it is triggered by the application of the Doppler shift $\epsilon_{DS}(\Phi)$ (flux bias $\Phi$) and/or by the phase difference $\phi$ between the superconductors. In the paper we have discussed in detail the case of asymmetric gaps which generalize previous studies of the same device to a more realistic situation where the right and left leads are not exactly identical. This is also necessary in order to investigate strong nonlinear conditions which typically occur in experiments. We have found that nonlocal thermoelectricity is present also in the asymmetric case. The nonlocal Onsager coefficients satisfy the standard symmetry between Seebeck and Peltier coefficients if the heat current between the two superconductors is properly defined to take into account the nonlocality. We have found that, in the asymmetric gap case, the Doppler shift develops a strong nonlocal thermoelectric effect when the two gaps close. We have discussed 
how the gap asymmetry gives rise to a new intermediate regime (for $\Delta_{L}<\abs{\epsilon_{DS}(\Phi)}<\Delta_{R}$) where Andreev interference determines a weak nonlocal thermoelectrical effect without applying a phase bias. We have investigated how this Andreev mechanism influences also the local thermoelectric effect.  
Furthermore, we have analyzed the heat current in the probe finding that it can be finite and the system can be viewed as a thermal router depending on the difference between the superconducting gaps.

Finally, we have discussed the nonlinear performance of the nonlocal thermoelectric machine.
In particular, we have studied the symmetry of the Seebeck coefficient, the power generated and the efficiency. 
The latter turns out to be quite low (with a maximum value $\eta/\eta_c\approx 3.5 \%$).
Notably, efficiency and power are maximized simultaneously for a wide range of parameters.
We have finally estimated that the Seebeck coefficient for large nonlinear temperature differences reaches a few tenths of $\mu$V/K, which is an impressive value given the operating temperature of few kelvin required by the BCS superconductors.  
We hope that this research will trigger further experimental investigations, since we expect that the stability against the possible gap asymmetries and the intensity of the effect would be detectable with nowadays low-temperature technologies.
\begin{acknowledgments}
 We acknowledge support from CNR-CONICET cooperation program “Energy conversion in quantum nanoscale hybrid devices”.  We are sponsored by PIP-RD 20141216-4905 
of CONICET, PICT-2017-2726 and PICT-2018-04536 from Argentina, as
well as the Alexander von Humboldt
Foundation, Germany (LA). A.B. and F.T acknowledge SNS-WIS joint lab QUANTRA. M. C. is supported by the Quant-Era project ``Supertop''. A. B. acknowledge the Royal Society through the International Exchanges between the UK and Italy (Grant No. IEC R2 192166).
\end{acknowledgments}

\appendix
\section{Scattering matrix}
\label{Scattering_matrix}

We analyze the setup assuming the absence of inelastic scattering since it may appear for junctions longer than coherence length $L\gg\xi=\hbar \varv_F/\pi\Delta$ and/or high temperatures where standard low temperature BCS superconductivity cannot survive. In such case dc transport is determined by the quantum mechanical scattering matrix $S$~\cite{Lambert98}, which yields scattering properties at energy $\epsilon$, of a phase-coherent, non-interacting system described by a Hamiltonian  ${\cal H}$ such as Eq.~(\ref{My_Hamiltonian}) .
The scattering problem, in terms of scattering matrix, can be formulated as
\begin{equation}
\label{scattering_equation}
\Psi^{\alpha}_{(i,\sigma)}|_{out}=S_{(i,\sigma)(j,\sigma')}^{\alpha\beta}\Psi^{\beta}_{(j,\sigma')}|_{in}
\end{equation}
It relates incoming/outgoing states $(j,\sigma')/(i,\sigma)$ with $\{\sigma,\sigma'\}=\{\uparrow,\downarrow\}$ labeling the spin-channel at the respective lead $i,j=N,L,R$. In Eq.~(\ref{scattering_equation}), $\{\alpha,\beta\}=\{e,h\}$ may indicate qps and qhs in the normal probe $N$, or eventually $\{\alpha,\beta\}=\{\tilde{e},\tilde{h}\}$ label QPs and QHs in the superconductors.
In order to compute the full scattering matrix $S$ of the system, we proceed by writing first the scattering matrix $S_N$ describing the coupling of the metallic probe with the TI edge 
\begin{equation}
\label{SN}
\mqty(c_{N  }^{{\color{black}\downarrow}}\\
b_{N  }^{{\color{black}\downarrow}}\\
c_{\rm{N} }^{{\color{black}\uparrow}}\\
b_{\rm{N}  }^{{\color{black}\uparrow}}\\
c_{L+}^{{\color{black}\downarrow}}\\
b_{L+}^{{\color{black}\downarrow}}\\
c_{\rm{R-} }^{{\color{black}\uparrow}}\\
b_{\rm{R-}  }^{{\color{black}\uparrow}})=
\begin{pmatrix}
\begin{matrix}
\begin{matrix}
0&0\\
0&0
\end{matrix}&\begin{matrix}
r&0\\
0&r^*
\end{matrix}\\
\begin{matrix}
r&0\\
0&r^*
\end{matrix}&\begin{matrix}
0&0\\
0&0
\end{matrix}
\end{matrix}&\begin{matrix}
\begin{matrix}
0&0\\
0&0
\end{matrix}&\begin{matrix}
t&0\\
0&t^*
\end{matrix}\\
\begin{matrix}
t&0\\
0&t^*
\end{matrix}&\begin{matrix}
0&0\\
0&0
\end{matrix}
\end{matrix}\\
\begin{matrix}
\begin{matrix}
0&0\\
0&0
\end{matrix}&\begin{matrix}
t&0\\
0&t^*
\end{matrix}\\
\begin{matrix}
t&0\\
0&t^*
\end{matrix}&\begin{matrix}
0&0\\
0&0
\end{matrix}
\end{matrix}&\begin{matrix}
\begin{matrix}
0&0\\
0&0
\end{matrix}&\begin{matrix}
r&0\\
0&r^*
\end{matrix}\\
\begin{matrix}
r&0\\
0&r^*
\end{matrix}&\begin{matrix}
0&0\\
0&0
\end{matrix}
\end{matrix}\\
\end{pmatrix}_{S_N}
\mqty(c_{N  }^{{\color{black}\uparrow}}\\
b_{N  }^{{\color{black}\uparrow}}\\
c_{\rm{N} }^{{\color{black}\downarrow}}\\
b_{\rm{N}  }^{{\color{black}\downarrow}}\\
c_{L+}^{{\color{black}\uparrow}}\\
b_{L+}^{{\color{black}\uparrow}}\\
c_{\rm{R-} }^{{\color{black}\downarrow}}\\
b_{\rm{R-}  }^{{\color{black}\downarrow}})
\end{equation}
where
we have indicated with $c_{i}^{\uparrow\downarrow}/\tilde{c}_{i}^{\uparrow\downarrow}$ and $b_{i}^{\uparrow\downarrow}/\tilde{b}_{i}^{\uparrow\downarrow}$ the incoming and outgoing electrons/qps and holes/qhs respectively with $i=L\pm,R\pm,N$ labeling the corresponding lead (with $-(+)$ indicating the left(right) side of the $S$-$TI$ interface). In particular, for the previous formula $S_N$, we assumed a symmetric beam splitter which effectively describes the contact interface between the normal lead $N$ and the TI where 
$r=\cos(\eta)$
and $t=i~\sin(\eta)$ with $\eta\in\left[0,\frac{\pi}{2}\right]$ such that unitarity is satisfied, i.e. $\abs{r}^2+\abs{t}^2=1$.
In the matrix the complex conjugate $r^*$ and $t^*$ are the amplitudes for the holes. Indeed $S_{N~(i,\sigma)(j,\sigma')}^{ee}(\epsilon)$ acts in the particle sector, and the scattering (sub-)matrix for the holes satisfies $S_{N~(i,\sigma)(j,\sigma')}^{hh}(\epsilon)=\left[S_{N~(i,\sigma)(j,\sigma')}^{ee}(-\epsilon)\right]^*$ as requested by the particle-hole symmetry of the system)~\cite{Pershoguba19}. Obviously the elements $S_{N~(i,\sigma)(j,\sigma')}^{\alpha\bar{\alpha}}(\epsilon)$ (with $\alpha=e,h$), coupling incoming electrons(holes) with outgoing holes(electrons) are necessarily zero since only ordinary scattering processes are involved, no Andreev reflections occur for normal metal probe.
The scattering matrix $S_N$ describing the contact between the normal-metal probe and the TI edge, can be also recasted in the more compact fashion $S_N(\eta)=i\sin{(\eta)}\zeta_1\otimes\sigma_1\otimes\tau_3+\cos{(\eta)}\zeta_0\otimes\sigma_1\otimes\tau_0$, expressed in terms of the Pauli matrices $\zeta,\sigma$ and $\tau$ respectively acting in the $N$-$TI$ channels space, spin space and PH space. 
Then we introduce the scattering matrices $S_{i}$ with $i=L/R$ describing respectively left/right interfaces of the TI with the superconductors. For the left interface we find

\begin{equation}
\label{SL}
\mqty(\tilde{c}_{L-}^{{\color{black}\downarrow}}\\
\tilde{b}_{L-}^{{\color{black}\downarrow}}\\
c_{\rm{L+} }^{{\color{black}\uparrow}}\\
b_{\rm{L+}  }^{{\color{black}\uparrow}})=
\begin{pmatrix}
\begin{matrix}
0&r^{L}_{\tilde{e}\tilde{h}}\\
r^{L}_{\tilde{h}\tilde{e}}&0
\end{matrix}&\begin{matrix}
t^{L}_{\tilde{e}e}&0\\
0&t^{L}_{\tilde{h}h}
\end{matrix}\\
\begin{matrix}
t^{L}_{e\tilde{e}}&0\\
0&t^{L}_{h\tilde{h}}
\end{matrix}&\begin{matrix}
0&r^{L}_{eh}\\
r^{L}_{he}&0
\end{matrix}
\end{pmatrix}_{S_{L}}
\mqty(\tilde{c}_{L-}^{{\color{black}\uparrow}}\\
\tilde{b}_{L-}^{{\color{black}\uparrow}}\\
c_{\rm{L+} }^{{\color{black}\downarrow}}\\
b_{\rm{L+}  }^{{\color{black}\downarrow}})
\end{equation}

that have been obtained solving the wave function matching problem at the specific interface between superconducting leads $L$ and the upper edge state of the TI. The obtained coefficients $r^{L}_{\alpha\beta}$ and $t^{L}_{\alpha\beta}$ represent respectively the reflection and transmission amplitudes of an incoming particle of type $\beta$ to a particle of type $\alpha$ at the interface. Those coefficients can be compactly written as
\begin{align}
r_{\gamma\bar{\gamma}}^{L}&=\gamma \frac{v_{L\bar{\gamma}}}{u_{L\bar{\gamma}}}e^{i\alpha_{\bar{\gamma}}}e^{i\gamma \phi_{L}}\nonumber\\
r_{\tilde{\gamma}\tilde{\bar{\gamma}}}^{L}&=-\frac{v_{L\gamma}}{u_{L\gamma}}e^{-i\beta_{\gamma}^{L}}{\color{black}\cdot \Xi_{\gamma}^{L} (\epsilon)}\nonumber\\
t_{\gamma\tilde{\gamma}}^{L}&=\frac{\sqrt{u_{L\bar{\gamma}}^2-v_{L\bar{\gamma}}^2}}{u_{L\bar{\gamma}}}e^{\frac{i}{2}\left(\alpha_{\bar{\gamma}}-\beta_{\bar{\gamma}}^{L}\right)}e^{-i\gamma\frac{\phi_{L}}{2}}e^{i\bar{\gamma} \lambda}\cdot \Xi_{\bar{\gamma}}^{L} (\epsilon)\nonumber \\
t_{\tilde{\gamma}\gamma}^{L}&=\bar{\gamma}\frac{\sqrt{u_{L\gamma}^2-v_{L\gamma}^2}}{u_{L\gamma}}e^{\frac{i}{2}\left(\alpha_{\gamma}-\beta_{\gamma}^{L}\right)}e^{i\bar{\gamma}\frac{\phi_{L}}{2}}e^{i\bar{\gamma} \lambda}\cdot \Xi_{\gamma}^{L} (\epsilon)
\label{scatcoeff}
\end{align}
where the qp/qh index ($\gamma=e/h$) in the \textit{l.h.s.} is converted in a simple sign ($\gamma=+/-$) in the \textit{r.h.s.} to match with the notation used in Eq.~(\ref{u_v}). We also introduced the symbol $\Xi_{\gamma}^{L}(\epsilon)=1$ when $\abs{\epsilon_{DS}}>\Delta_{L}$ $\wedge$ $0<\epsilon<\abs{\Delta_{L}-\abs{\epsilon_{DS}}}$ and $\Xi_{\gamma}^{L}(\epsilon)=\Theta (\abs{\epsilon_{\gamma}}-\Delta_{L})$ otherwise. 
The exponents take the values
$\alpha_{\gamma}=\epsilon_{\gamma}/\epsilon_c$ and $\beta_{\gamma}^{i}=\sqrt{\epsilon_{\gamma}^2-\Delta_{i}^2}/\epsilon_c$
with $\epsilon_c=\frac{\hbar \varv_F}{L}$ the Thouless energy of the junction. The phase $\lambda=2\arg{\left(1+i \frac{\Lambda_L}{2\hbar \varv_F}\right)}$ accounts of the contact potential $\Lambda_L(x)$ in the BdG Hamiltonian of Eq.~(\ref{My_Hamiltonian}).
A similar result for the scattering matrix at the right interface $S_{R}$ can be computed. The scattering coefficients can be obtained from Eq.~(\ref{scatcoeff}) by replacing $(r_{\alpha\beta}^{L},t_{\alpha\beta}^{L})\rightarrow(r_{\beta\alpha}^{R},t_{\beta\alpha}^{R})$ in the \textit{l.h.s.} and, in the \textit{r.h.s.}, making the substitution $L\to R$ and $\phi_{L}\to -\phi_{R}$.

In conclusion, following the standard procedure presented in Ref.~\onlinecite{Datta95}, the full scattering matrix of the system is obtained by combining the three scattering matrices such as
\begin{equation}
\label{S_comb}
S=S_{L}\circ S_{N}\circ S_{R}
\end{equation} 
which determines the fundamental scattering between the three terminals.

\section{Symmetries}
\label{Symmetries}
It is known~\cite{Lambert98,Jacquod12,Blasi_2020_Andreev_Inteferometer} that the scattering coefficients satisfy relations due to microreversibility $P_{i,j}^{\alpha,\beta}(\epsilon,\vec{\theta})=P_{j,i}^{\beta,\alpha}(\epsilon,-\vec{\theta}),$
particle-hole symmetry $P_{i,j}^{\alpha,\beta}(\epsilon,\vec{\theta})=P_{i,j}^{-\alpha,-\beta}(-\epsilon,\vec{\theta})$ and unitarity
\begin{align}
\sum_{\alpha i}P_{i,j}^{\alpha,\beta}(\epsilon,\vec{\theta})=N_j^{\beta}(\epsilon,\vec{\theta}))\nonumber\\
\sum_{\beta j}P_{i,j}^{\alpha,\beta}(\epsilon,\vec{\theta})=N_i^{\alpha}(\epsilon,\vec{\theta}),\nonumber
\end{align}
where $N_i^{\alpha}(\epsilon,\vec{\theta})$ is the number of open channels for $\alpha$-like QPs at energy $\epsilon$ in lead $i$.
Moreover, when $\Delta_{L}=\Delta_{R}$ (namely, the system is left/right symmetric), the scattering coefficients show also the following additional symmetries
\begin{align}
\label{P_properties}
P_{N,N}^{\alpha,\beta}(\epsilon,\vec{\theta})&=P_{N,N}^{-\alpha,-\beta}(\epsilon,\vec{\theta}),\nonumber\\
P_{N,L}^{\alpha,\beta}(\epsilon,\vec{\theta})&=P_{N,R}^{-\alpha,-\beta}(\epsilon,\vec{\theta}),\nonumber\\
P_{N,{L/R}}^{\alpha,\beta}(\epsilon,\vec{\theta})&=P_{N,{L/R}}^{-\alpha,-\beta}(\epsilon,-\vec{\theta}).
\end{align}
Interestingly, again in the left/right symmetric case, one finds that the scattering coefficients $P_{i,j}^{\alpha,\beta}$ do not depend on the position $x_0$ of the probe, but simply on the total length $L$ of the junction.
The reason for this relies on the symmetry exhibited by the different paths that take a QP of type $\beta$ from lead $j$ to a QP of type $\alpha$ in lead $i$.
More specifically, due to the helicity of the edge state and the spin-independence of the transmission amplitude $t$, each of these paths comes in pair with its symmetric one (obtained 
by exchanging left and right), in such a way that their contribution to $P_{i,j}^{\alpha,\beta}$ only depends on $L$.
On the contrary, in the asymmetric case, i.~e. when $\Delta_{\rm L}\neq\Delta_{\rm R}$, the position of the probe $x_0$ along the edge does matter. When we discuss the asymmetric case usually we assume to fix the probe tip just in the middle of the junction, i.e. $x_0=L/2$.

\section{Analytical results of the probe's currents in the symmetric case $\Delta_{L}=\Delta_{R}$}
\label{appendix_Symmetric_case}

\begin{figure*}[ht]
\centering
   \includegraphics[width=1\textwidth]{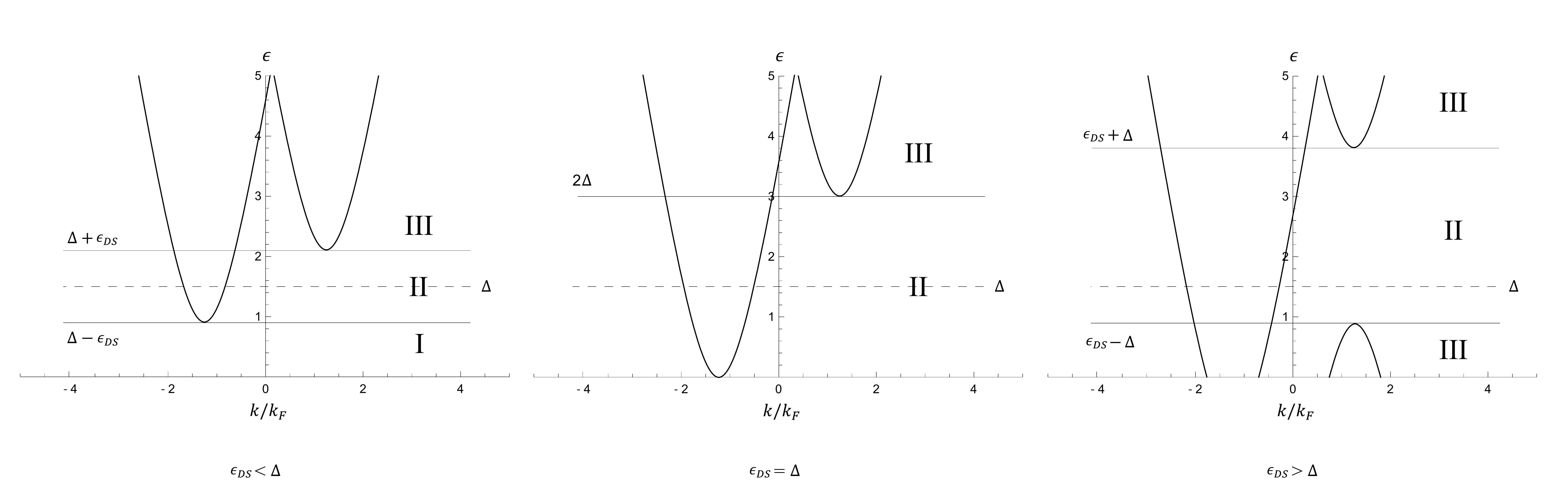}
    \caption{Regions of validity in Eq.~(\ref{AQ_analytical}) for $\epsilon_{DS}>0$: $\epsilon_{DS}<\Delta$ (left panel), $\epsilon_{DS}=\Delta$ (middle panel), $\epsilon_{DS}>\Delta$ (right panel).} 
    \label{Regions_I_II_III} 
\end{figure*}

Here we discuss some analytical results for the symmetric case (i.~e.~$\Delta_L=\Delta_R=\Delta$) exploiting the symmetry expressed in the relations of Eqs.~(\ref{P_properties}). We concentrate mainly on the quasi-particle charge ($\gamma=-$) and heat ($\gamma=+$) current at the probe $N$ that can be written as

\begin{widetext}

\begin{equation}
\label{J00}
J_N^\gamma=\frac{2}{h}\int_{0}^{\infty}d\epsilon~e^{\delta_{\gamma ,-}}(\epsilon-eV_N)^{\delta_{\gamma ,+}}\Big\{F_N^\gamma(\epsilon)\left[ A^\gamma(\epsilon,\vec{\theta})+ A^\gamma(\epsilon,-\vec{\theta})\right]- F_{LR}^\gamma(\epsilon)\left[Q^\gamma(\epsilon,\vec{\theta})+\gamma Q^{\gamma}(\epsilon,-\vec{\theta})\right]\Big\}
\end{equation}

\end{widetext}

in which we defined the Fermi function sums/differences ($\gamma=+/-$) at the probe $F_N^{\gamma}= f_N^+ +\gamma f_N^-$ or among the two superconductors $F_{LR}^{\gamma}= f_{\rm{L}}^{\pm}+\gamma f_{\rm{R}}^{\mp}$. The quantity
 \begin{align}
\label{A}
A^\gamma (\epsilon,\vec{\theta})&=\left(N_N^+-P^{++}_{NN}+\gamma P^{-+}_{NN}\right)/2\nonumber\\
&=\left(N_N^--P^{--}_{NN}+\gamma P^{+-}_{NN}\right)/2
\end{align}
represents the strength of the qp charge (with $\gamma=-$) or the heat flux (with $\gamma=+$) transferred from the probe into the edge at given energy $\epsilon$ and $\vec{\theta}$. The scattering probabilities $P^{\pm\pm}_{NN}$ describe normal reflections, $P^{\pm\mp}_{NN}$ the Andreev ones and $N_N^{+(-)}$ the number of open channels for electrons (holes) at the probe. The quantity
\begin{align}
\label{Q}
Q^\gamma (\epsilon,\vec{\theta})&=\left(P^{++}_{NS_L}+\gamma P^{-+}_{NS_L}\right)\nonumber\\
&=(-1)^{\delta_{\gamma ,-}}\left(P^{+-}_{NS_R}+\gamma P^{--}_{NS_R}\right).
\end{align}
describes the strength of the charge (heat) $\gamma=-$ ($\gamma=+$)  transferred into the probe $N$ when a qp is injected from the superconductor $L$. For $\gamma=-$, due to the gap symmetry and Eq.~(\ref{P_properties}), it coincides with amount of qh (with opposite sign) transferred into the probe when a qh is injected from $R$.
Similarly, when $\gamma=+$, Eq.~(\ref{Q}) represents the amount of energy transferred into the probe when a qp(qh) is injected from $L$($R$).

These quantities have been discussed in Ref.~\onlinecite{Blasi_2020_Andreev_Inteferometer} in the case of the charge current (i.~e.~$\gamma=-$) and without the Doppler shift $\Phi=0$. 
Here we generalized them also to the presence of the Doppler shift (namely $\vec{\theta}\neq 0$), in which case their analytical expressions read as:
\begin{widetext}

\begin{empheq}[left={A^\gamma (\epsilon,\vec{\theta})=\quad\empheqlbrace\,}]{equation}
\begin{split}
&\frac{2\abs{t}^4\cdot\delta_{\gamma,-}}{1+\abs{r}^4+2\abs{r}^2\cos{(2\pi\frac{ L\epsilon_{+}}{\xi\Delta}+\phi_{u}-2\arccos{(\frac{\epsilon_{+}}{\Delta})})}}\nonumber\\
& \frac{(g(\epsilon)^2-\gamma)(g(\epsilon)^2+\gamma\abs{r}^2)\abs{t}^2\cdot \Theta(\epsilon_{DS})}{g(\epsilon)^4+\abs{r}^4-2g(\epsilon)^2\abs{r}^2\cos{(2\pi\frac{L\epsilon_{+}}{\xi\Delta} +\phi_u)}}+\frac{2\abs{t}^4\cdot \Theta(\epsilon_{DS})~\delta_{\gamma,-}}{1+\abs{r}^4+2\abs{r}^2\cos{(2\pi\frac{ L\epsilon_{+}}{\xi\Delta}-\phi_{u}-2\arccos{(\frac{\epsilon_{+}}{\Delta})})}}
\nonumber\\
&\frac{(g(\epsilon)^2-\gamma)(g(\epsilon)^2+\gamma\abs{r}^2)\abs{t}^2}{g(\epsilon)^4+\abs{r}^4-2g(\epsilon)^2\abs{r}^2\cos{(2\pi\frac{L\epsilon_{+}}{\xi\Delta}+\phi_u)}}
\end{split}
\quad\quad
\begin{split}
&\text{for } \epsilon\in I\nonumber\\ \\ 
&\text{for } \epsilon\in II\nonumber\\ \\
&\text{for } \epsilon\in III
\end{split}
\end{empheq}
\begin{empheq}[left={Q^\gamma (\epsilon,\vec{\theta})=\quad\empheqlbrace\,}]{equation}
\begin{split}
&0\\
&\frac{(g(\epsilon)^2-1)(g(\epsilon)^2+\gamma\abs{r}^2)\abs{t}^2}{g(\epsilon)^4+\abs{r}^4-2g(\epsilon)^2\abs{r}^2\cos{(2\pi\frac{L\epsilon_{+}}{\xi\Delta}+ \phi_u)}}\cdot \Theta(\epsilon_{DS})
\\
&\frac{(g(\epsilon)^2-1)(g(\epsilon)^2+\gamma\abs{r}^2)\abs{t}^2}{g(\epsilon)^4+\abs{r}^4-2g(\epsilon)^2\abs{r}^2\cos{(2\pi\frac{L\epsilon_{+}}{\xi\Delta}+ \phi_u)}}
\end{split}
\quad\quad
\begin{split}
&\text{for } \epsilon\in I\\ \\
&\text{for } \epsilon\in II\\ \\
&\text{for } \epsilon\in III
\end{split}
\label{AQ_analytical}
\end{empheq}

\end{widetext}
where  $g(\epsilon)=e^{\arccosh(\epsilon_{+}/\Delta)}$, $\epsilon_{+}=\epsilon+\epsilon_{DS}$, $\abs{r}^2=1-\abs{t}^2$ and $\phi_u=\phi+\frac{2L\epsilon_{DS}}{\pi\xi\Delta}$ is the phase difference along the edge which includes the contribution of the external magnetic flux. In the expressions of Eq.~(\ref{AQ_analytical}), we indicated the energy regions $I$, $II$ and $III$, depicted in Fig.~\ref{Regions_I_II_III}, which represent, respectively, the contributions deriving from the sub-gap (region I),  the semi-continuum (region II) and the full continuum (regions III).
We notice that in Eq.~(\ref{AQ_analytical}) the sub-gap contribution of the function $A^\gamma(\epsilon,\vec{\theta})$ is nonzero only in the case of charge current (for $\gamma=-$), while it is zero for the heat current (for $\gamma=+$). This is because the Andreev bound states cannot allow any thermal transport, while mediating only the charge transport. 

From the previous analytical formulas we can deduce some general consequences for the probe's currents of Eq.~(\ref{J00}).
Let us consider first the case of the charge current with $\gamma=-$. When $V_N=0$, the function $F^-_N=0$ (since $f_N^+=f_N^-$), so from formula Eq.~(\ref{J00}) one can easily conclude that $J_N^-$ is independent of the temperature $T_N$.
This shows that no \emph{local} thermoelectrical effect can be induced by means of a thermal bias between the
TI and the probe. The only thermoelectric response in the probe is the \emph{nonlocal} one when a thermal bias between the two superconductors $\delta T$ is applied, i.e. $F_{LR}^{-}= f_{\rm{L}}^{\pm}- f_{\rm{R}}^{\mp}\neq0$. This is particularly important at experimental level since the temperature of the
probe does not need to be controlled during the measurement of nonlocal thermoelectricity. 
The strength of such nonlocal thermoelectric response (see Eq. (\ref{J00})) is determined by the integral over the energies of the odd parity component in $\vec{\theta}$ of the function $Q^- (\epsilon,\vec{\theta})$, i.~e.~$Q^-(\epsilon,\vec{\theta})- Q^{-}(\epsilon,-\vec{\theta})$, from which it follows that the Onsager nonlocal thermoelectrical linear coefficient $L_{12}$ is an odd function of $\vec{\theta}$.
Instead, when $\delta T=0$, the function $F_{LR}^-=0$ (since $f_L^\pm=f_R^\mp$), so from formula Eq. (\ref{J00}) it turns out that the charge current $J_N^-$ is determined by the integral over the energies of the even parity component in $\vec{\theta}$ of the function $A^- (\epsilon,\vec{\theta})$, i.~e.~$A^-(\epsilon,\vec{\theta})+ A^{-}(\epsilon,-\vec{\theta})$, from which it follows that the Onsager local electrical coefficient $L_{11}$ is an even function of $\vec{\theta}$

Furthermore, regarding the heat component of Eq.(\ref{J00}) with $\gamma=+$, it turns out that $J_N^+$ is an even function of $\vec{\theta}$ since it depends only on the even components of the functions $A^+ (\epsilon,\vec{\theta})$ and $Q^+ (\epsilon,\vec{\theta})$. In the linear regime one can find that, for $T_{L/R}=T \pm \delta T/2$ and $T_N=T$, the heat current in the probe is proportional to the energy integral of the term
$\sum_{\sigma=\pm}\left[A^+(\epsilon,\sigma\vec{\theta})-Q^+(\epsilon,\sigma\vec{\theta})\right]$, which - using the expressions of Eq.(\ref{AQ_analytical}) - turns out to be zero. 
As a consequence, in the linear regime, the heat current at the probe $J_N^+=0$.
This is a direct consequence of the energy conservation computed in the linear regime and for symmetric gaps.

\end{document}